\documentclass[twocolumn,english,aps,prl,superscriptaddress,longbibliography]{revtex4-2}
\usepackage[T1]{fontenc}
\usepackage[latin9]{inputenc}
\usepackage{amsmath}
\usepackage{ dsfont }
\usepackage{amssymb}
\usepackage{mathdots}
\usepackage{graphicx}

\makeatletter
\usepackage{epsfig}
\usepackage{dcolumn}
\usepackage{bm}
\usepackage{bbm}
\usepackage{amsfonts}
\usepackage{latexsym}
\usepackage[normalem]{ulem}

\usepackage{lipsum}
\usepackage{hyperref}
\hypersetup{colorlinks=true, pdfstartview=FitV, linkcolor=blue, citecolor=blue, plainpages=false, pdfpagelabels=true, urlcolor=blue}
\usepackage[all]{hypcap}

\usepackage{babel}

\makeatother

\usepackage{babel}

\newcommand{\snte}{Sn$_{1-x}$Pb$_{x}$Te$_{1-y}$Se$_{y}$}

\begin{document}
\title{Corner states, hinge states and Majorana modes in SnTe nanowires}
\author{Nguyen Minh Nguyen}
\affiliation{International Research Centre MagTop, Institute of Physics, Polish
Academy of Sciences, Aleja Lotnikow 32/46, PL-02668 Warsaw, Poland}
 \author{Wojciech Brzezicki}
\affiliation{International Research Centre MagTop, Institute of Physics, Polish Academy of Sciences,  Aleja Lotnikow 32/46, PL-02668 Warsaw, Poland} 
\affiliation{Institute of Theoretical Physics, Jagiellonian University, ulica S. \L{}ojasiewicza 11, PL-30348 Krak\'ow, Poland}
 \author{Timo Hyart}

\affiliation{International Research Centre MagTop, Institute of Physics, Polish Academy of Sciences, Aleja Lotnikow 32/46, PL-02668 Warsaw, Poland}
\affiliation{Department of Applied Physics, Aalto University, 00076 Aalto, Espoo, Finland}

\begin{abstract}
SnTe materials are one of the most flexible material platforms for exploring the interplay of topology and different types of symmetry breaking.   
We study symmetry-protected topological states in 
SnTe nanowires in the presence of various combinations of Zeeman field, 
$s-$wave superconductivity and inversion-symmetry-breaking field. We uncover the origin of robust corner states and hinge states in the normal state. In the presence of superconductivity, we find inversion-symmetry-protected gapless bulk Majorana modes, which give rise to quantized thermal conductance in ballistic wires. By introducing an inversion-symmetry-breaking field, the bulk Majorana modes become gapped and topologically protected localized Majorana zero modes appear at the ends of the wire.
\end{abstract}
\maketitle

\section{Introduction}

SnTe materials (\snte)  have already established themselves as paradigmatic systems for studying 3D topological crystalline insulators and topological phase transitions, because the band inversion can be controlled with the Sn content \cite{Hsieh2012, Story2012,Tanaka2012,Hasan2012}, but they may have a much bigger role to play in the future investigations of topological effects. Robust 1D modes were experimentally observed at the surface atomic steps  \cite{Sessi2016} and interpreted as topological flat bands using a model obeying a chiral symmetry \cite{Buczko2018}. Furthermore, the experiments  indicate that these flat bands may lead to correlated states and an appearance of a robust zero-bias peak in the tunneling conductance at low temperatures \cite{Mazur19}.  Although, there has been a temptation to interpret the zero-bias anomaly as an evidence of topological superconductivity, all the observed phenomenology can be explained without superconductivity or Majorana modes \cite{Mazur19, Brzezicki2019, timowojtek}. On the other hand, the standard picture of competing phases in flat-band systems \cite{Black-Schaffer, Ojajarvi} indicates that in thin films of SnTe materials  the tunability of the density with gate voltages may allow for the realization of both magnetism and superconductivity at the step defects. 
  
The improvement of the fabrication of  low-dimensional SnTe systems with a controllable carrier density has become increasingly pressing also because the theoretical calculations indicate that thin SnTe multilayer systems would support a plethora of 2D topological phases, including quantum spin Hall \cite{Liu2015, Safaei2015} and 2D topological crystalline insulator phases \cite{Liu2014, Brzezicki2019}. The realization of 2D topological crystalline insulator phase would be particularly interesting, because in these systems a tunable breaking of the mirror symmetries would open a path for new device functionalities  \cite{Liu2014}. Indeed, the recent experiments in thin films of SnTe materials indicate that the transport properties of these systems can be controlled by intentionally breaking the mirror symmetry \cite{kazakov2020dephasing}.   Furthermore, SnTe materials are promising candidate systems for studying the  higher order topology \cite{Schindlereaat0346,Hsu18}. Thus, it is an outstanding challenge to develop approaches for probing the hinge and corner states in these systems.

In addition to the rich topological properties, SnTe materials are also one of the most flexible platforms for studying the interplay of various types of symmetry breaking fields. The superconductivity can  be induced via the proximity effect or by In-doping, and both theory and experiments indicate rich physics emerging as a consequence \cite{PhysRevLett.109.217004, PhysRevB.87.140507, Bernevig14, APL17, adma, Bliesener19, PhysRevB.100.241402, trimble2020josephson}. Interesting topological phases are predicted to arise also in the presence of a Zeeman field \cite{Bernevig14b}, which breaks the time-reversal symmetry. In experiments, the Zeeman field can be efficiently applied with the help of external magnetic field by utilizing the huge $g$ factor  $g \sim 50$ \cite{Dybko, leadsalts}, or it can be introduced with the help of magnetic dopants \cite{Story86, Story90, PhysRevB.85.045210}. 
While the magnetism and superconductivity are part of the standard toolbox for designing topologically nontrivial phases, the SnTe materials offer also unique opportunities for controlling the topological properties by breaking the crystalline symmetries. In particular, it is possible to break the inversion symmetry by utilizing ferroelectricity or a structure inversion asymmetry \cite{Chang2016, Kim2019, ZFu2019, Valentine17, Lee2020, Rafal21}. This has already enabled the realization of a giant Rashba effect, and it may be important also for the topology.  

So far the topological properties of SnTe nanowires have received little attention experimentally, but this may soon change due to the continuous progress in their fabrication  \cite{doi:10.1021/acsaelm.0c00740, Sadowski}. In this paper, we systematically study the symmetries and topological invariants in SnTe nanowires and propose to utilize the tunable symmetry-breaking fields for realizing different types of topological states. After describing the system (Sec.~\ref{sec:HamAndSym}), we consider Zeeman field parallel to the wire and show that depending on the Zeeman field magnitude and the wire thickness there exists four qualitatively different behaviors around the charge neutrality point: trivial insulator regime, one-dimensional Weyl semimetal phase, band-inverted insulator regime and indirect semimetal phase (Sec.~\ref{sec:Zeeman}).
We show that the band-inverted insulator regime is characterized by a pseudospin texture and an appearance of low-energy states localized at the corners of the wire, whereas the Weyl semimetal phase is protected by a non-symmorphic screw-axis rotation symmetry ($4$-fold rotational symmetry) in the case of even (odd) thicknesses, and the low-energy states are localized at the hinges of the wire. We uncover how these hinge states are related to the topological corner states appearing in two-dimensional  Hamiltonians belonging to the Altland-Zirnbauer class DIII \cite{PhysRevB.55.1142} in the presence of a rotoinversion symmetry, and we explicitly construct an analytical formula for a $\mathbb{Z}_2$ topological invariant describing their existence (Sec.~\ref{sec:hing}).  
In the presence of superconductivity (Sec.~\ref{Sec:Supercond}), we find inversion-symmetry-protected gapless topological bulk Majorana modes, which give rise to quantized thermal conductance in ballistic wires.  Finally, we show that by introducing an inversion-symmetry-breaking field, the bulk Majorana modes become gapped and topologically protected localized Majorana zero modes appear at the ends of the wire.

\section{Hamiltonian and its symmetries \label{sec:HamAndSym}}

\begin{figure}
\includegraphics[width=0.9\columnwidth]{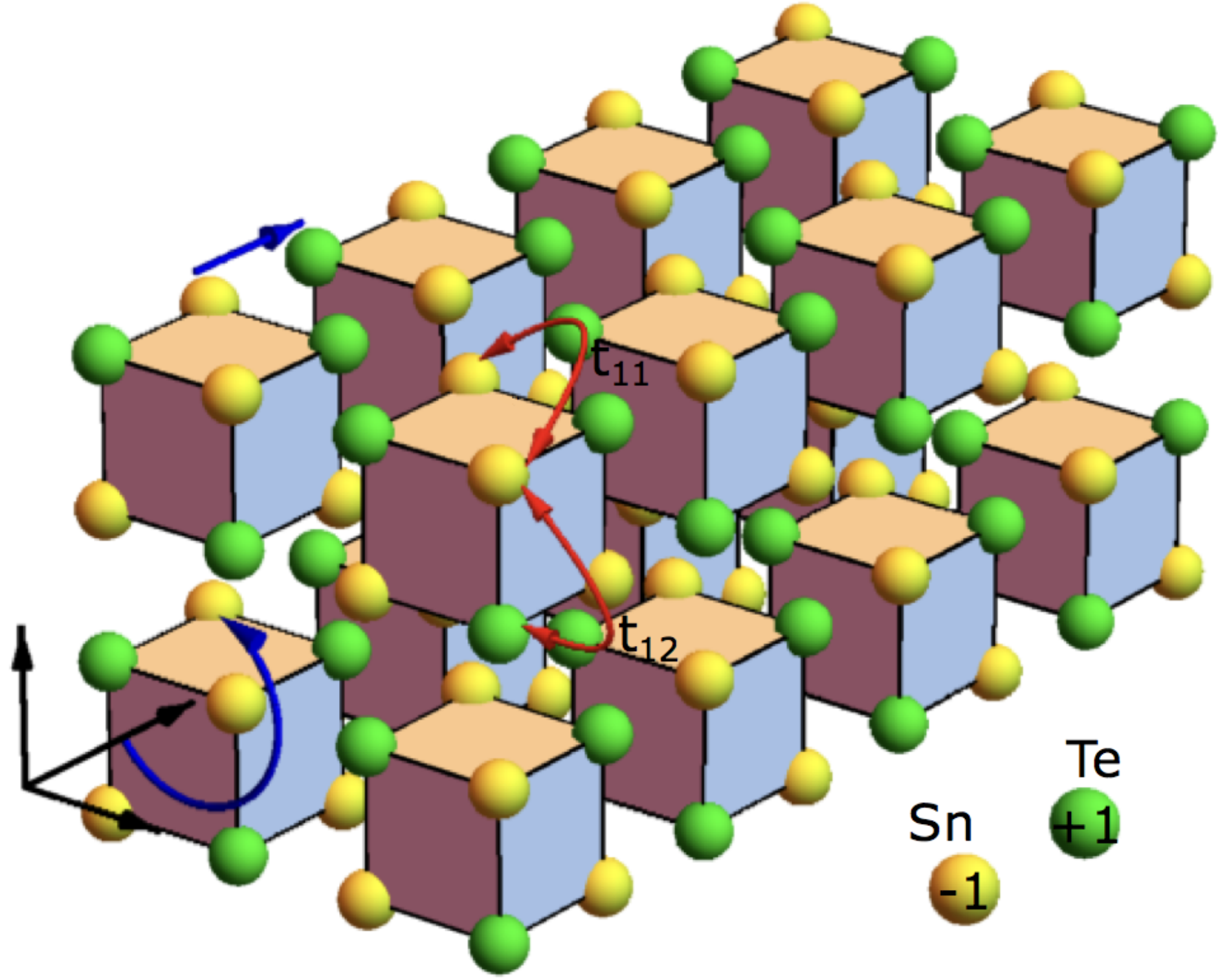}
\caption{Schematic view of the system. The blue arrow indicates the screw-axis operation, including a $\pi/2$ rotation with respect to z axis and a half-lattice vector translation. The red arrows depict the nearest-neighbor ($t_{12}$) and the next-nearest-neighbor ($t_{11}$) hopping terms in Hamiltonian (\ref{Ham}). The two sublattices, corresponding to Sn and Te atoms, have opposite onsite energies.  \label{fig:1} }
\end{figure}

Our starting is the $p$-orbital tight-binding Hamiltonian  
\begin{eqnarray}
\mathcal{H}(\textbf{k}) &=&  m \mathds{1}_2\!\otimes\!\mathds{1}_3\!\otimes\!\Sigma + t_{12}\!\!\!\sum_{\alpha=x,y,z}\!\!\mathds{1}_2\!\otimes\!(\mathds{1}_3-\mathit{L}_{\alpha}^{2} )\!\otimes\!h_{\alpha} (k_{\alpha} ) \nonumber \\
&&\hspace{-0.8cm}+t_{11} \! \sum_{\alpha \neq \beta} \! \mathds{1}_2  \!\otimes\! \left[ \mathds{1}_3 - \dfrac{1}{2}\left(\mathit{L}_{\alpha}+ \epsilon_{\alpha\beta}\mathit{L}_{\beta}\right)^2\right] \!\otimes\! h_{\alpha,\beta} (k_{\alpha},k_{\beta}) \Sigma  \nonumber \\
&&\hspace{-0.8cm}+ \!\! \sum_{\alpha=x,y,z} \!\! \lambda_{\alpha}\,\sigma_{\alpha} \!\otimes\! \mathit{L}_{\alpha} \!\otimes\! \mathds{1}_8,
\label{Ham}
\end{eqnarray}
which has been  used for describing the bulk topological crystalline insulator phase in the SnTe materials \cite{Hsieh2012} and various topological phases in lower dimensional systems \cite{Sessi2016, Brzezicki2019}. Here we have chosen a cubic unit cell containing eight lattice sites  (Fig.~\ref{fig:1}), $\Sigma$ is a diagonal $8 \times 8$ matrix with entries $\Sigma_{i,i}=\mp1$ at the two sublattices (Sn and Te atoms), $\varepsilon_{\alpha\beta}$ is Levi-Civita symbol, $L_{\alpha}=-i \varepsilon_{\alpha \beta \gamma}$ are the $3 \times 3$ angular momentum $L=1$ matrices, $\sigma_{\alpha}$ are Pauli matrices, and $h_{\alpha}(k_{\alpha})$ and $h_{\alpha,\beta}(k_{\alpha},k_{\beta} )$ are $8 \times 8$ matrices describing  hopping between the nearest-neighbors and next-nearest-neighbor sites, respectively (see Appendix \ref{app:nanowire}).  In investigations of topological properties it is useful to allow the spin-orbit coupling to be anisotropic, hence $\lambda_{\alpha}$, although the reference physical case is $\lambda_{\alpha}\equiv\lambda$. When not otherwise stated we use  $m = 1.65$ eV, $t_{12}=0.9$ eV, $t_{11}=0.5$ eV  and $\lambda =0.3$ eV. 

We first consider an infinite nanowire along the $z$-direction with $N_x=N_y$ unit cells in $x$ and $y$ directions. The Hamiltonian for the nanowire $\mathcal{H}^{\rm 1D} (k_{z} )$ can be constructed using Hamiltonian (\ref{Ham})  and it satisfies a fourfold screw-axis symmetry  (see Appendix \ref{app:nanowire})
\begin{equation}
S_c\left(k_{z} \right) = \mathit{P}_z\otimes e^{-i \frac{\pi}{4}\mathcal{\sigma}_{z}} \otimes e^{-i \frac{\pi}{2}\mathit{L}_{z}}\otimes  s_c \left( k_z \right),
\label{eqSc} 
\end{equation} 
where $\mathit{P}_z$ and  $s_c\left(k_{z} \right)$ realize a transformation of the lattice sites, consisting of a  translation by a half lattice vector and $\pi/2$  rotation with respect to the $z$ axis (Fig.~\ref{fig:1}), between the unit cells  and inside the unit cell, respectively (see Appendix \ref{app:nanowire}).
Additionally, there exists also glide plane symmetries $M_x(k_z)$ ($M_y(k_z)$) consisting of a mirror reflection with respect to $\hat{x}$ ($\hat{y}$) plane and a half-lattice translation along $z$ axis and diagonal mirror symmetries $M_{xy}$ ($M_{yx}$) with respect to the $\hat{x}+\hat{y}$ ($\hat{x}-\hat{y}$) planes. The product of $M_x(k_z)$ with $M_y(k_z)$ or $M_{xy}$ with $M_{yx}$ yields a twofold rotation symmetry with respect to the $z$ axis. All these symmetry act at any $k_z$ can be used to block-diagonalize the 1D Hamiltonian. 
The mirror symmetry $M_z(k_z)$ with respect to the $z$ plane acts on the Hamiltonian as $M_z(k_z)\mathcal{H}^{\rm 1D}(k_{z})M_z(k_z)^{\dagger}=\mathcal{H}^{\rm 1D}(-k_{z})$ and the inversion symmetry operator can be constructed as $I\propto M_x(k_z)M_y(k_z)M_z(k_z)$. 

If the wire has odd number of atoms in $x$ and $y$ directions, it cannot be constructed from full unit cells, and this  influences the symmetries of the system. In particular,
for odd thicknesses the screw-axis rotation symmetry is replaced by an ordinary $4$-fold rotation symmetry. 

\section{Topological states in the presence of Zeeman field \label{sec:Zeeman}}

\begin{figure}[b]
\hspace{-0.8cm}
\includegraphics[width=0.95\columnwidth]{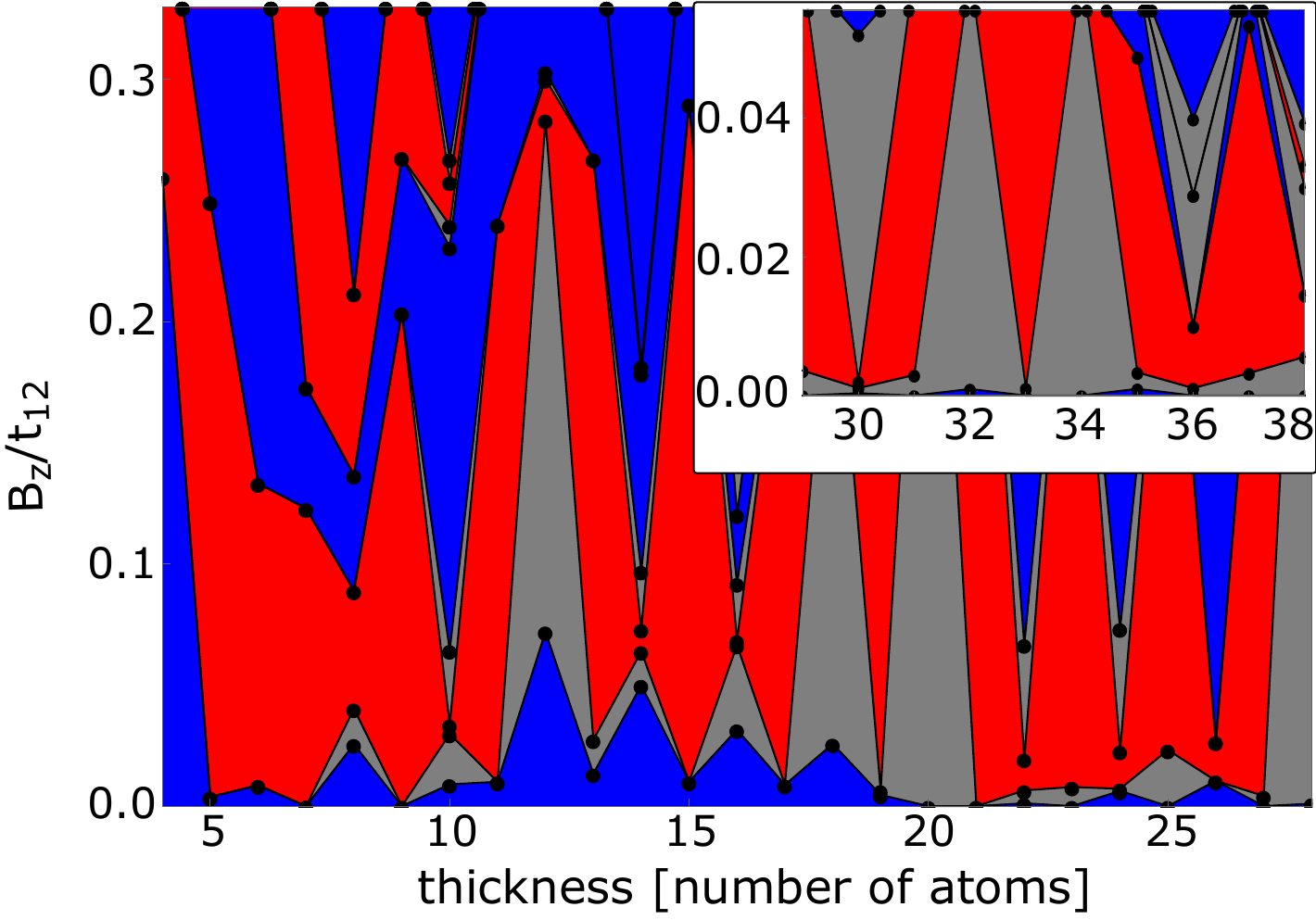}
\vspace{-0.2cm} 
\caption{Phase diagram as function of the nanowire thickness (dimensions of the square cross-section) and Zeeman  field $B_z$. The different phases are: insulator phase (blue), Weyl semimetal phase (red) and indirect semimetal phase (grey). In the regime of large Zeeman field the insulator phase supports a pseudospin texture due to band inversion, resulting in the appearance of localized corner states  (see Fig.~\ref{fig:5}). Dots show the actually computed phase boundaries at discrete values of the wire thickness. 
\label{fig:2}}
\end{figure}

\begin{figure}
\includegraphics[width=1\columnwidth]{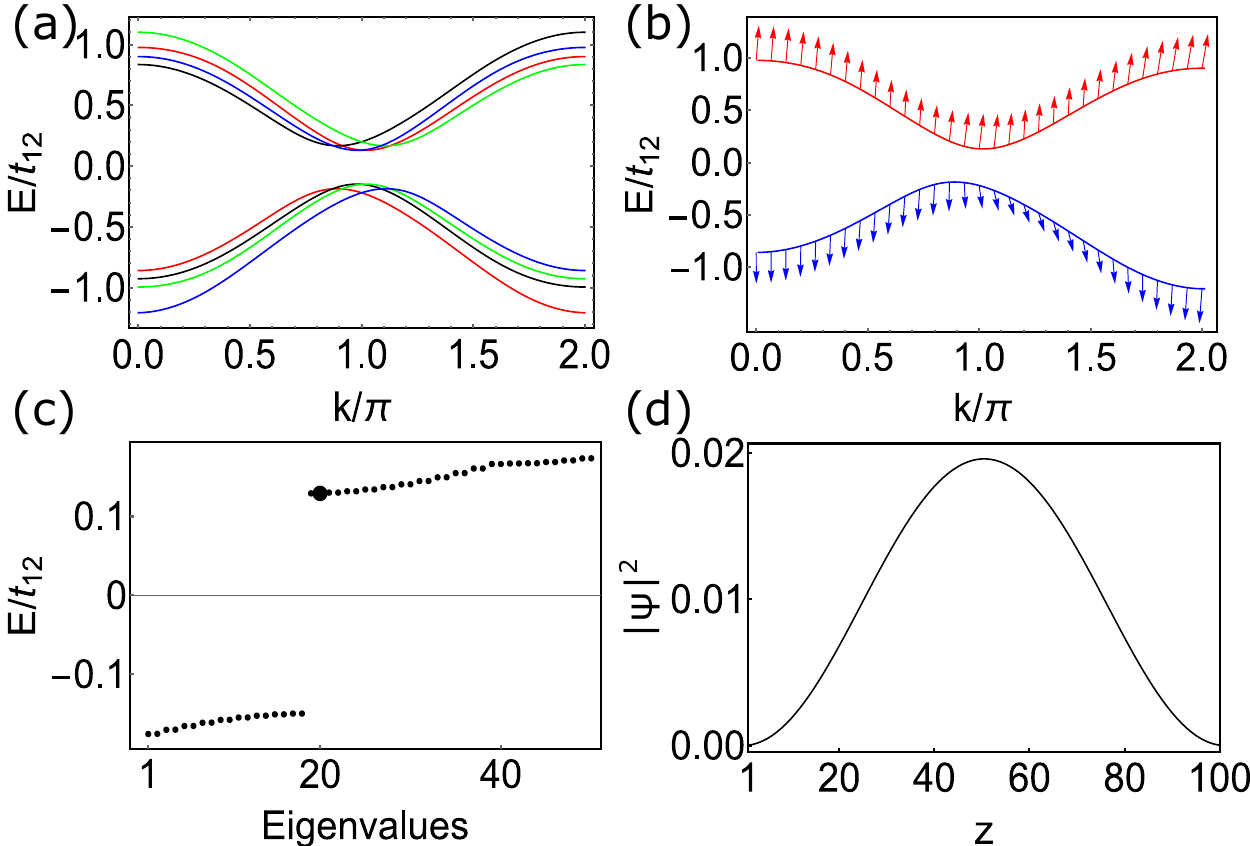}\caption{ (a) Low-energy band structure in the trivial insulator phase. The different colors indicate the various $S_c$ eigenvalue subspaces. (b) The sublattice pseudospin  texture for a pair of bands with the same $S_c$ eigenvalue. The texture is in-plane because $\langle\tau_y\rangle$ is negligible. (c) Energy spectrum of 200 atoms long wire showing no end-states. (d) LDOS for the bulk state highlighted in plot (c).  
The thickness of the wire is 4 atoms and $B_z = 0.1 t_{12}$.
\label{fig:3}}
\end{figure}

\begin{figure}
\includegraphics[width=1\columnwidth]{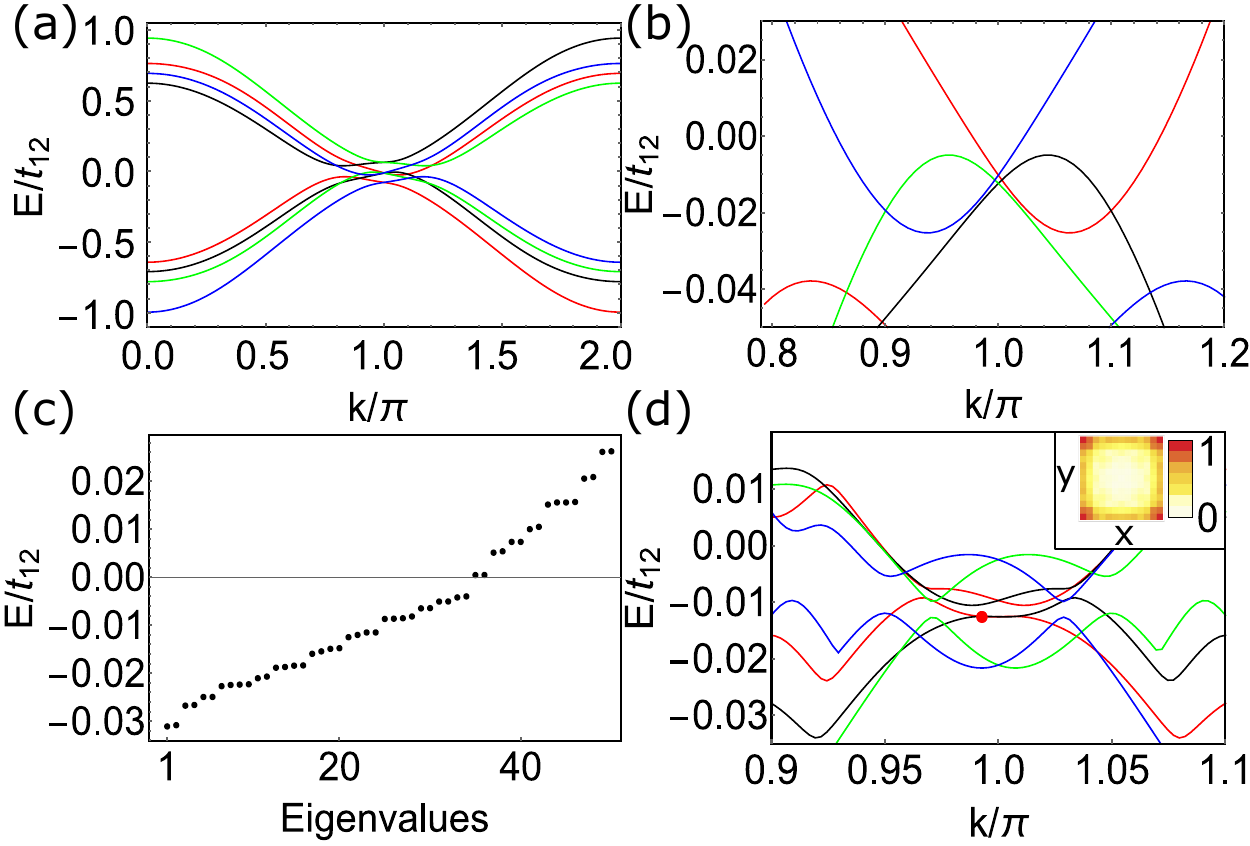}
\caption{
(a) Low-energy band structure in the Weyl semimetal phase for  4 atoms thick nanowire and $B_z = 0.3 t_{12}$. The different colors indicate various $S_c$ eigenvalue subspaces. (b) Band crossings around $k_z=\pi$. (c)  Energy spectrum of 200 atoms long wire.
(d) Band structure of 26 atoms thick nanowire at $B_z = 0.012 t_{12}$. At the band-crossing point (red dot) the states are localized in the hinges of the wire. Inset: LDOS (normalized with the maximum value) projected to the square cross section of the wire.  
\label{fig:4}}
\end{figure}

\begin{figure}
\includegraphics[width=1\columnwidth]{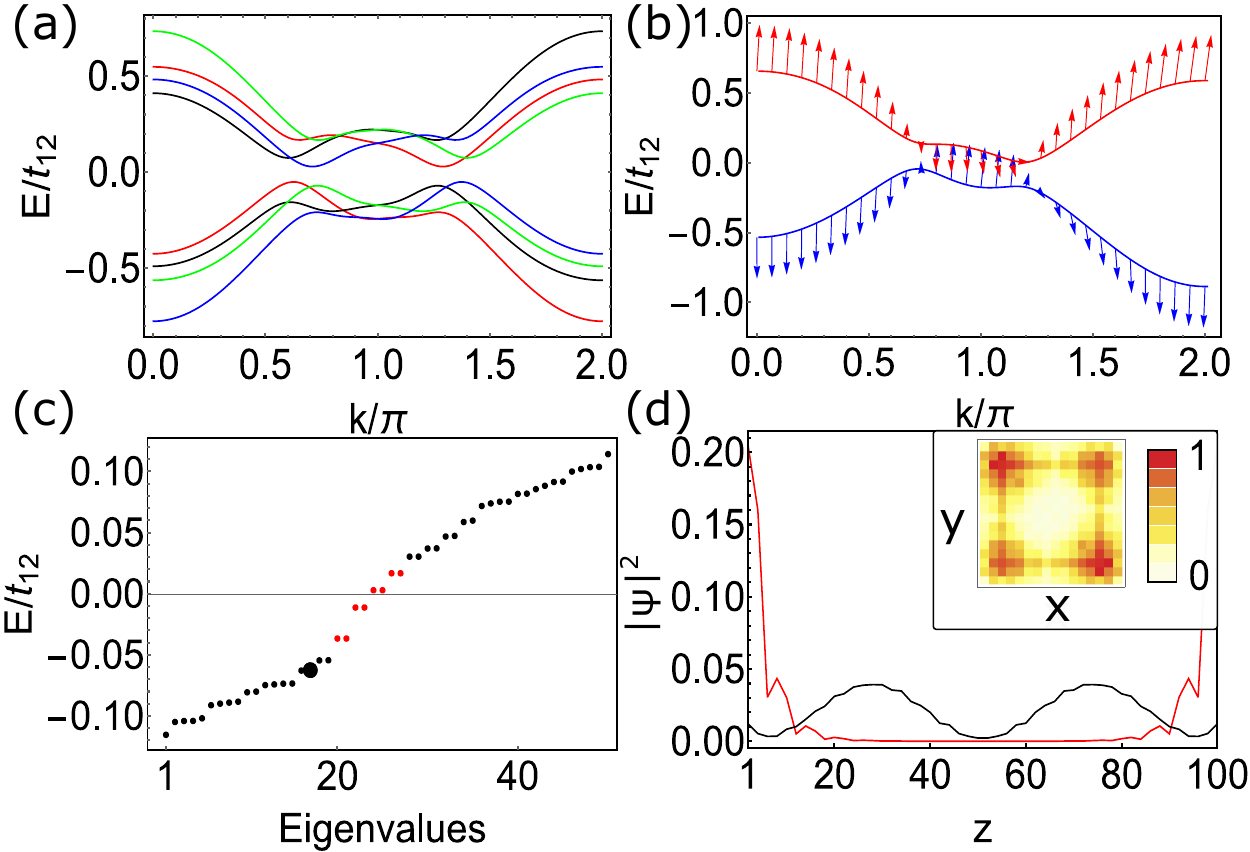}
\caption{
(a) Low-energy band structure in the band-inverted insulator phase for the 4 atoms thick nanowire and $B_z = 0.4 t_{12}$. (b) The sublattice pseudospin  texture for a pair of bands with the same $S_c$ eigenvalue. (c) Energy spectrum for 200 atoms long wire. The red points indicate eight corner states and black points are bulk states. (d) LDOS as a function of $z$ for the corner states (red line) and a bulk state (black line). Inset: LDOS (normalized with the maximum value) as a function of $x$ and $y$ for the corner states in a 6 atoms thick nanowire. 
\label{fig:5}}
\end{figure}

In this section, we study the properties of the system in the presence of Zeeman field ${\cal H}_Z=\mathbf{B} \cdot \vec{\sigma}$ applied along the wire $\mathbf{B}=(0,0,B_z)$. This field breaks the time-reversal symmetry $T$ and the mirror symmetries $M_x(k_z)$, $M_y(k_z)$, $M_{xy}$  and $M_{yx}$, but it preserves the inversion symmetry $I$, the mirror symmetry $M_z(k_z)$ and  the screw-axis symmetry  $S_c\left(k_{z} \right)$.  
We find that as a function of Zeeman field magnitude and the wire thickness there exists four qualitatively different behaviors around the charge neutrality point: trivial insulator regime, one-dimensional Weyl semimetal phase, band-inverted insulator regime and indirect semimetal phase (Fig.~\ref{fig:2}). The differences between these phases are summarized in Figs.~\ref{fig:3}, \ref{fig:4} and \ref{fig:5}.


In the case of small Zeeman field $B_z$ we find a trivial insulating phase or an indirect semimetal phase depending on the wire thickness. Neither of these phases supports in-gap states localized at the ends of the wire. By increasing $B_z$ we find that there appears a  Weyl semimetal phase for a range of wire thicknesses and Zeeman field magnitudes (Figs.~\ref{fig:2} and \ref{fig:4}). For even thicknesses of the wire the band crossings (Weyl points) are protected by the non-symmorphic screw-axis rotation symmetry $S_c(k_z)$, which allows us to decompose the Hamiltonian into 4 diagonal blocks, so that the energies of eigenstates belonging to different blocks (indicated with different colors in Fig.~\ref{fig:4}) can cross. Due to this reason a change in the number of eigenstates belonging to specific eigenvalues of the screw-axis rotation symmetry below the Fermi level as a function of $k_z$ can be used as a topological invariant for the Weyl semimetal phase. In the case of odd thicknesses the screw-axis rotation symmetry is replaced by an ordinary $4$-fold rotational symmetry, but also this symmetry can be utilized to block diagonalize the Hamiltonian at any $k_z$, and therefore it can protect the Weyl semimetal phase in a analogous way. 
  Note that due to non-symmorphic character of $S_c(k_z)$  the $S_c$-subblocks of ${\cal H}^{\rm 1D}(k_z)$ are $4\pi$-periodic and they transform into each other in $2\pi$ rotations. Because symmetries guarantee that the spectrum is symmetric around $k_z=\pi$, this elongated period leads to forced band crossings at $k_z=\pi$, see Figs.~\ref{fig:3}-\ref{fig:5}(a). 
However,  the forced band crossings typically appear away from the zero energy (Figs.~\ref{fig:3} and \ref{fig:5}) and therefore they do not guarantee the existence of Weyl semimetal phase at the charge neutrality point. In the Weyl semimetal phase shown in Fig.~\ref{fig:4} the crossings occur between an electron-like band and a hole-like band carrying different $S_c$ eigenvalues so that there necessarily exists states at all energies. As demonstrated by calculating the local density of states (LDOS)  in Fig.~\ref{fig:4}(d)  the states connecting the conduction and valence bands are often found to be localized at the hinges of the nanowire. We discuss the origin of these hinge states in Sec.~\ref{sec:hing}.

By increasing $B_z$ further we find another insulating phase for a wide range of wire thicknesses and Zeeman field magnitudes. In this case, the band dispersions have the camel's back shape [Fig.~\ref{fig:5}(a)] which typically appears in topologically nontrivial materials, but we have checked that these band structures can be adiabatically connected to the trivial insulator phase, and therefore the topological nature may only be related to an approximate symmetry of the system. Nevertheless, the bands  support a nontrivial pseudospin texture $\langle {\vec \tau} \rangle_p=\langle \psi_p(k)|\vec{\tau}|\psi_p(k)\rangle$, where the pseudospin operators $\tau_{\alpha}$ are the Pauli matrices acting in the sublattice space. In the high-field insulating phase the pseudospin component $\langle \tau_y \rangle_p$ is negligible and the pseudospin direction rotates in two-dimensional  $(\langle \tau_x \rangle_p, \langle \tau_z \rangle_p)$-space so that its direction is inverted around $k_z=\pi$ [see Fig.~\ref{fig:5}(b)], whereas in the low-field insulator phase the sublattice pseudospin texture is trivial [Fig.~\ref{fig:3}(b)]. Therefore we call the insulating phases as band-inverted and trivial insulators, respectively. The band-inverted insulator phase also supports subgap end states localized at the corners of the wire [Fig.~\ref{fig:5}(c),(d)], whereas no subgap end states can be found in the trivial insulator phase [Fig.~\ref{fig:3}(c),(d)]. 

We emphasize that the thin nanowires are used here only for illustration purposes because in these cases the strengths of the Zeeman fields required for realizing the different behaviors of the system are not experimentally feasible. However, with increasing thickness of the nanowires the Weyl semimetal and band-inverted phases occur at smaller values of $B_z$, so that in the case of a realistic thickness they can be accessed with feasible magnitudes of the Zeeman field. The inset of Fig.~\ref{fig:2} shows a zoom into the experimentally most relevant regime of the phase diagram. 

The Weyl points are protected by the screw-axis symmetry (or $4$-fold rotation symmetry) which is broken if the Zeeman field is rotated away from the $z$-axis. However, if we utilize an approximation  $\lambda_{z}=0$ and $\lambda_{x}=\lambda_{y}=\lambda$, there exists also a non-symmorphic chiral symmetry $S_z(k_z)$, and it is possible to combine it with $M_z(k_z)$  and time-reversal $T$ to construct an antiunitary operator that anticommutes with $\mathcal{H}^{\rm 1D}(k_{z})$ for any $k_z$. This operator squares to $+1$ and gives rise to a Pfaffian-protected Weyl semimetal phase also when the Zeeman field is not along the $z$-axis (Appendix \ref{sec:lz0}). We also point out that if the screw-axis symmetry (or 4-fold rotational symmetry) is broken so that the system supports only a 2-fold rotational symmetry (e.g. due to anisotropic spin-orbit coupling or rectangular nanowires with $N_x \ne N_y$), the 2-fold rotational symmetry can still protect the existence of the Weyl points (Appendix \ref{sec:anisotropy}).

\section{Hinge states \label{sec:hing}}

\begin{figure}[t]
\includegraphics[width=1\columnwidth]{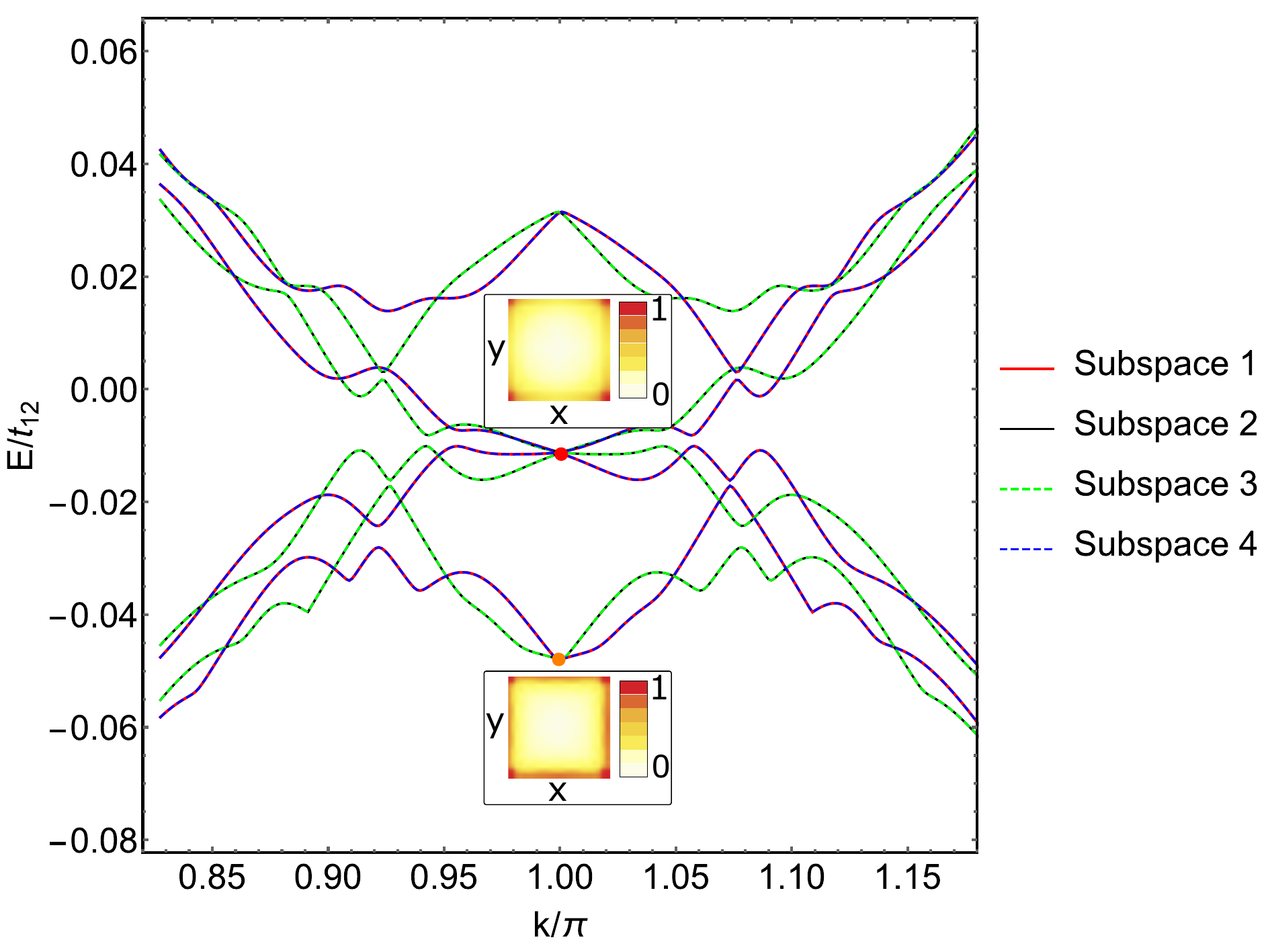}
\caption{
Low-energy band structure and the LDOS (normalized with the maximum value) demonstrating the existence of hinge states in the absence of Zeeman field $\mathbf{B}=0$  for 50 atoms thick nanowire. 
\label{fig:8}}
\end{figure}

\begin{figure}[t]
\includegraphics[width=0.6\columnwidth]{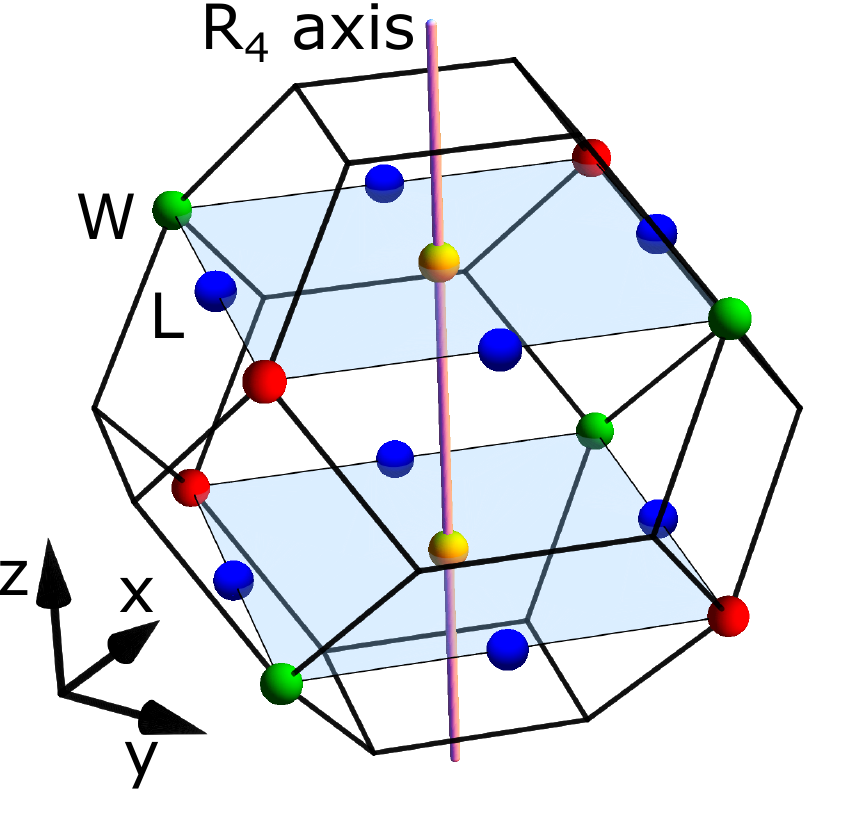}\caption{
Brillouin zone of the rock-salt crystals. The shaded planes form a $k_3=\pi$ plane. The blue balls ($L$ points) are the high-symmetry points $k_{1,2}=0,\pi$
within the $k_3=\pi$ plane. The yellow balls are the fourfold rotation $R_4$ centers at $(k_1,k_2)=(\pi,\pi)/2$ and $(k_1,k_2)=(3\pi,3\pi)/2$. The red and green balls are the 
rotoinversion centers ($W$ points) at $(k_1,k_2)=(\pi,3\pi)/2$ and $(k_1,k_2)=(3\pi,\pi)/2$, respectively.
\label{fig:12}}
\end{figure}

\begin{figure}
\includegraphics[width=1\columnwidth]{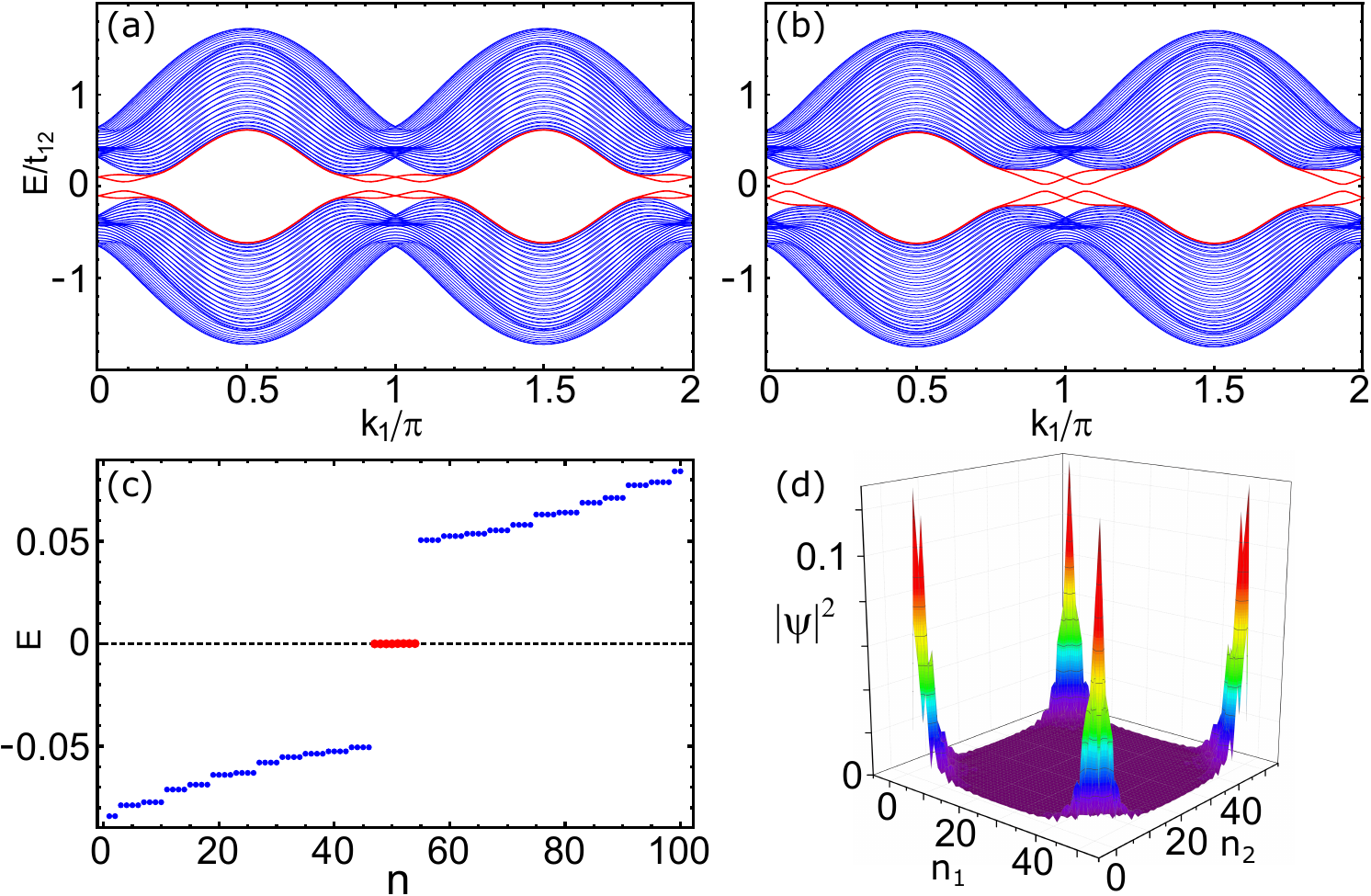}\caption{ 
(a) Low-energy spectrum of the Hamiltonian ${\cal H}(k_1,k_2,k_3=\pi)$ for  
$\lambda_z=0$ in the case of open boundary conditions in the $a_2$ direction. The width of the system is $50$ unit cells.
(b) The low-energy spectrum in the case of uniform spin-orbit coupling $\lambda_{\alpha}=\lambda$ ($\alpha=x,y,z$).
(c)  Spectrum for $\lambda_z=0$ in the case of open boundary conditions in both $a_1$ and $a_2$ directions. The red dots show the 8 zero-energy corner states. The system size is $50\times 50$ unit cells. 
(d) LDOS of the corner states. Here, $n_1$ and $n_2$ label the unit cells in the directions of $a_1$ and $a_2$, respectively.
\label{fig:9}}
\end{figure}

We find that in addition to the Weyl semimetal phase at $B_z \ne 0$ [Fig.~\ref{fig:4}(d)], the hinge states appear also in the absence of Zeeman field [Fig.~\ref{fig:8}], and they resemble the protected states appearing in higher-order topological phases \cite{Schindlereaat0346,Trifunovic19, Geier20, Trifunovic21}. SnTe materials have been acknowledged as promising candidates for higher-order topological insulators  but the gapless surface Dirac cones appearing at the mirror-symmetric surfaces make the experimental realization difficult \cite{Schindlereaat0346}. This problem can be avoided if the system supports a 2D higher-order topological invariant for a specific high-symmetry plane in the ${\bf k}$ space where the surface states are gapped - this plane is shown in Fig.~\ref{fig:12}. To explore this possibility,  we study the bulk Hamiltonian ${\cal H}(k_1,k_2,k_3)$ describing a system with 
inequivalent atoms at positions $(0,0,\pm 1/2)$  and lattice vectors $a_1=(1,0,1)$, $a_2=(0,1,1)$ and $a_3=(0,0,2)$ (see Appendix \ref{sec:2uc}). We find  that the 2D Hamiltonian ${\cal H}(k_1,k_2,\pi)$ with $\mathbf{B}=0$ supports edge states [red lines in Fig.~\ref{fig:9}(a),(b)], but a small energy gap is opened due to $\lambda_x = \lambda_y=\lambda$ spin-orbit coupling terms. 
The spectrum is similar both for  $\lambda_z =0$ [Fig.~\ref{fig:9}(a)] and $\lambda_z=\lambda$ [Fig.~\ref{fig:9}(b)] so that neglecting $\lambda_z$ is a good approximation.
Moreover, the numerics indicates that two adjacent edges of the system are topologically distinct leading to appearance of zero-energy corner states at their intersection [Figs.~\ref{fig:9}(c),(d)].

We find that the presence of the corner states is described by a $\mathbb{Z}_2$ topological invariant. To construct the invariant, we note that ${\cal H}(k_1,k_2,\pi)$ with $\lambda_z =0$ obeys a chiral symmetry $S_z = \sigma_z\otimes\mathbbm{1}_3\otimes\tau_x$, where $\sigma_z$ refers to spin, $\mathbbm{1}_3$ to 
orbitals and $\tau_x$ to sublattice degrees of freedom. The Hamiltonian also obeys the time-reversal
symmetry $T=\sigma_y\otimes\mathbbm{1}_3\otimes\mathbbm{1}_2$, which anticommutes with $S_z$,  so that the 
Hamiltonian belongs to class DIII. In the eigenbasis of 
$S_z$ the Hamiltonian and time-reversal operator have block-off-diagonal forms 
\begin{equation}
{\cal H}(k_{1},k_{2},\pi)=\begin{pmatrix}0 & u(k_{1},k_{2})\\
u^{\dagger}(k_{1},k_{2}) & 0
\end{pmatrix}
\end{equation}
and
\begin{equation}
T=\begin{pmatrix}0 & -i\mathbbm{1}_6\\
 i\mathbbm{1}_6 & 0
\end{pmatrix}.
\end{equation}
Thus, $u(k_1,k_2)^T=-u(-k_1,-k_2)$ and therefore we can define a Pfaffian at the time-reversal invariant points ${\bf K}$
\begin{equation}
p={\rm Pf}\,u(\boldsymbol{K})
\end{equation}
By utilizing inversion symmery $I = \mathbbm{1}_2\otimes\mathbbm{1}_3\otimes\tau_z$  we get that 
$p$ is a real number (see Appendix \ref{sec:pf}). In our model $p$ can be evaluated explicitly and it takes the form
\begin{equation}
p=(m-4t_{11})(m+2t_{11})^2-2m\lambda^2.
\end{equation}
Notice that $p$ is the same for all time-reversal invariant points ${\bf K}$ in the ($k_1$,$k_2$)-plane  due to the symmetries of the model (see Appendix \ref{sec:2uc}). 
In the usual notation of the 3D Brillouin zone of the rock-salt crystals, the time-reversal invariant points ${\bf K}$ in the ($k_1$,$k_2$)-plane correspond to the $L$ points (see Fig.~\ref{fig:12}).
 
Interestingly, we find that $p$ does not give a complete description of the presence of the corner states, because we also need to consider the high-symmetry point $\boldsymbol{K'}=(\pi/2,3\pi/2)$. This point is a rotoinversion center, so that in the eigenbasis of the rotoinversion operator the Hamiltonian takes a block-diagonal form 
\begin{equation}
{\cal H}(\boldsymbol{K'})=\begin{pmatrix}h_1 & 0 & 0 & 0\\
0 & h_2 & 0 & 0\\
0 & 0 & h_3 & 0\\
0 & 0 & 0 & h_4
\end{pmatrix}.
\end{equation}
By utilizing the chiral symmetry, inversion symmetry and time-reversal symmetry we find that (see Appendix \ref{sec:det})
\begin{equation}
{\rm det}[{\cal H}(\boldsymbol{K'})] =d^4,
\end{equation}
where $d \equiv \det h_1=\det h_4 = - \det h_2 = - \det h_3$, 
and therefore $d$ changes sign at the zero-energy gap closing occurring at the momentum $\boldsymbol{K'}$. 
In our model $d$ can be evaluated analytically and it takes the form
\begin{equation}
d=\det v_1=m((m-2t_{11})^2+4t_{12}^2)-2(m-2t_{11})\lambda^2.
\end{equation}
In the usual notation of the 3D Brillouin zone of the rock-salt crystals, the rotoinversion centers $\boldsymbol{K'}$ are  the $W$ points (see Fig.~\ref{fig:12}).

\begin{figure}
\includegraphics[width=0.7\columnwidth]{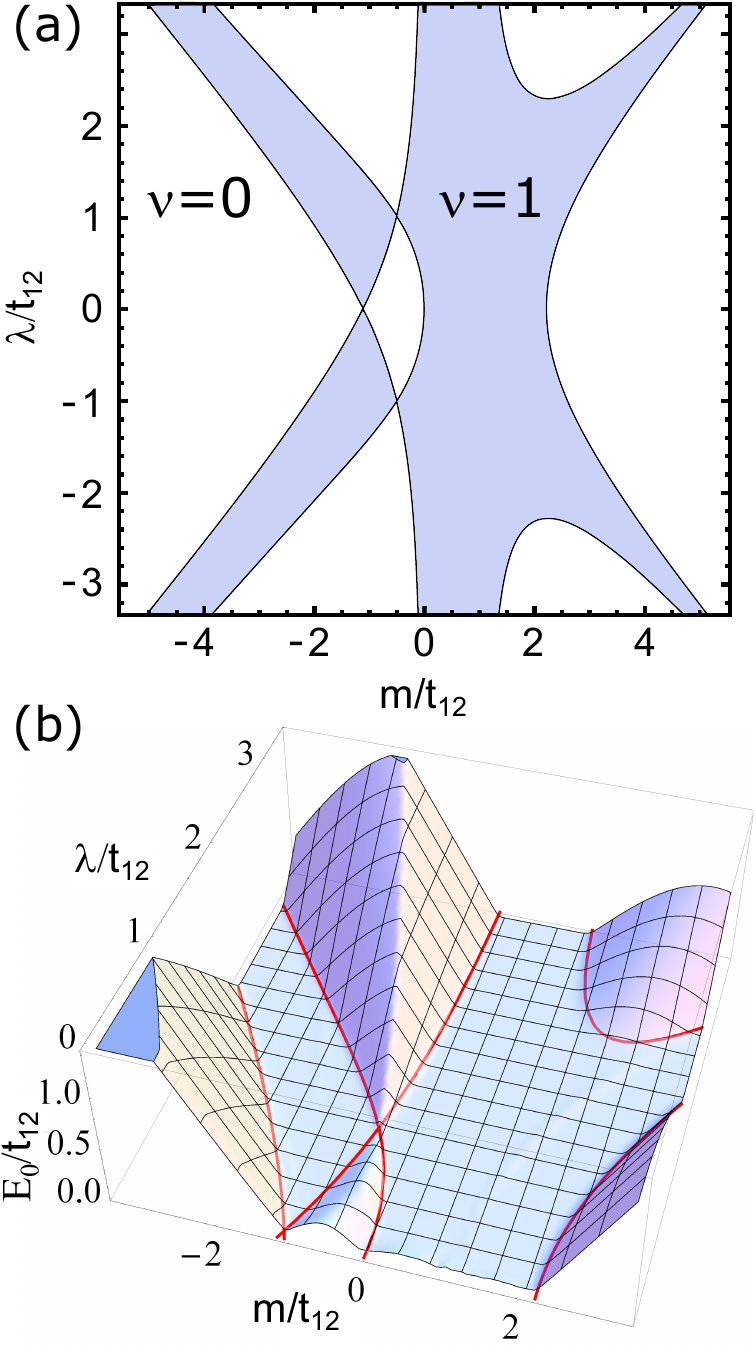}\caption{
(a) Topological invariant $\nu$ [Eq.~(\ref{topinv-hinge})] of the 2D Hamiltonian ${\cal H}(k_1,k_2,\pi)$ in the chiral limit  
$\lambda_z = 0$ and $\lambda_x = \lambda_y = \lambda$ as function of $m$ and $\lambda$. The shaded region is the non-trivial phase $\nu=1$ supporting corner states. (b) Energy of a state being closest to the zero energy in the system with all edges open, as function of $m$ and $\lambda $. The red lines are phase boundaries from (a). The system size is $50\times 50$ unit cells
\label{fig:10}}
\end{figure}

\begin{figure}
\includegraphics[width=0.7\columnwidth]{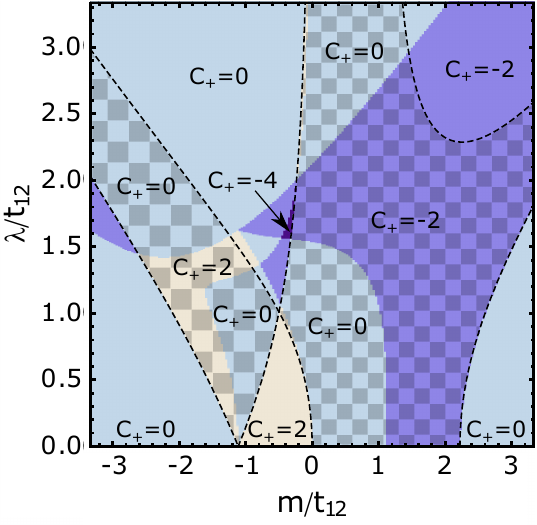}\caption{
Topological phase diagram of Hamiltonian ${\cal H}(k_1,k_2,k_3)$ in the $m$---$\lambda$ plane.
Colors indicate different mirror $M_{xy}$ Chern numbers $C_+$ defined in the $k_1$---$k_3$ plane (with $k_2=k_1$).  
Areas bounded by the dashed line and filled with checkerboard pattern are non-trivial in the sense of 
$\nu$ invariant of Eq.~(\ref{topinv-hinge}), also shown in Fig.~\ref{fig:10}. The chiral limit is assumed with 
$\lambda_z = 0$ and $\lambda_x = \lambda_y = \lambda$.
\label{fig:11}}
\end{figure}

The  $\mathbb{Z}_2$ invariant $\nu$ can be determined using $d$ and $p$ as
\begin{equation}
\nu = (1-{\rm sgn}(pd))/2.
\label{topinv-hinge}
\end{equation}
The topological phase diagram in the $m$---$\lambda$ plane is given in Fig.~\ref{fig:10}(a). By comparing to the corner state spectrum shown in Fig.~\ref{fig:10}(b), we find that $\nu$ describes the appearance of the corner states perfectly in our model. In the nontrivial phase $\nu=1$, there are two localized states at every corner. They are Kramers partners and carry opposite chirality eigenvalues. We emphasized that the topological invariant (\ref{topinv-hinge}) is not directly related to topological crystalline insulator invariant of the SnTe materials (see Fig. \ref{fig:11}). Therefore, we expect that it is possible to find material compositions supporting the higher-order topological phase outside the topological crystalline insulator phase and vice versa.

\section{Majorana modes in the presence of superconductivity \label{Sec:Supercond}}

Majorana zero modes are intensively-searched non-Abelian quasiparticles which hold a promise for topological quantum computing \cite{Nay08,Sar15, Beenakker20}. The key ingredients for realizing Majorana zero modes are usually thought to be spin-orbit coupling (spin-rotation-symmetry breaking field), magnetic field (time-reversal-symmetry breaking field) and superconductivity \cite{flensberg2021engineered}, and there exists a number of candidate platforms for studying Majorana zero modes including chains of adatoms \cite{Choy11, Nad14, Kimme16}  and various strong spin-orbit coupling materials in the presence of  superconductivity and magnetism \cite{Fu08, Lut10, Oreg10, Mou12}.  SnTe materials are particularly promising candidates for this purpose because in addition to the strong spin-orbit coupling they offer flexibility for introducing symmetry breaking fields such as superconductivity, magnetism and inversion-symmetry breaking fields.

In the presence of induced $s-$wave superconductivity the Bogoliubov-de Gennes Hamiltonian for the nanowires has the form 
\begin{equation}
H^{sc}(k_z)=\begin{pmatrix}{\cal H}^{1D}(k_z)-\mu & i\sigma^{y}\Delta\\
-i\sigma^{y}\Delta & -({\cal H}^{1D}(-k_z)^{T}-\mu) 
\end{pmatrix}, \label{SC-Ham}
\end{equation}
where $\sigma_y$ acts in the spin space. This Hamiltonian obeys a particle-hole symmetry 
\begin{equation}
C^{sc} H^{sc}(-k_z)^T C^{sc}=-H^{sc}(k_z),
\end{equation}
where
\begin{equation}
C^{sc}=\begin{pmatrix}0 & \mathds{1}\\
\mathds{1} & 0
\end{pmatrix}.
\end{equation}
We can utilize $C^{sc}$ to perform a unitary transformation on the Hamiltonian so that in the new basis the Hamiltonian $H_{U}^{sc}(k_z)$ is antisymmetric at $k_z=0, \pi$ and  ${\rm Pf} H^{sc}_U(k_z=0, \pi) \in \mathbb{R}$ (see Appendix \ref{sec:strongsc}).
Since $iH_{U}^{sc}(0,\pi)\in \mathbb{R}$ we use real Schur decomposition to evaluate the Pfaffian in a numerically
stable way, as suggested in \cite{Wimmer11}.
Therefore, we can define a $\mathbb{Z}_2$ topological invariant as 
\begin{equation}
\nu_{sc}=\left(1-{\rm sgn}[{\rm Pf} H^{sc}_U(0)  {\rm Pf}  H^{sc}_U(\pi)] \right)/2.
\end{equation}
This is the strong topological invariant of 1D superconductors belonging to the class D. In fully gapped 1D superconductors the $\nu_{sc}=1$ phase supports unpaired Majorana zero modes localized at the end of the wire.

The Hamiltonian (\ref{SC-Ham}) also satisfies an inversion symmetry
\begin{equation}
I^{sc}(k_z)=\begin{pmatrix}I(k_z) & 0\\
0 & I(k_z)
\end{pmatrix}.
\end{equation}
The product of $C^{sc}$ and $I^{sc}(k_z)$ is an antiunitary chiral operator, which allows to perform another unitary transformation on the Hamiltonian, so that  in the new basis the Hamiltonian $H_{V}^{sc}(k_z)$ is antisymmetric at all values of $k_z$ and  ${\rm Pf} H^{sc}_V(k_z) \in \mathbb{R}$ (see Appendix \ref{sec:inv-bulkMajo}). Therefore, consistently with classification of gapless topological phases \cite{Zhao16}, we can define an inversion-symmetry protected $\mathbb{Z}_2$ topological invariant for all values of $k_z$ as
\begin{equation}
\nu_I (k_z) = \left(1- {\rm sgn}[{\rm Pf} H_{V}^{sc}(k_z)] \right)/2.
\end{equation}
If this invariant changes as a function of $k_z$ there must necessarily be a gap closing. Thus, it enables the possibility of a topological phase supporting inversion-symmetry protected gapless bulk Majorana modes. Here, we use the term Majorana in the same way as it is standardly used in the physics literature, such as Refs. \cite{Cham10,*Bee14}, so that it can be used to refer to all Bogoliubov quasiparticles in superconductors. In the presence of inversion symmetry the protection of the gapless Majorana bulk modes is similar to the Weyl points in Weyl semimetals: They can only be destroyed by merging them in a pairwise manner. 
The experimental signature of the Majorana bulk modes in ballistic wires is quantized thermal conductance $G_{\rm th}=(G_0/2) \sum_n T_n$ in units of thermal conductance quantum $G_0=\pi^2 k_B^2 T/(3h)$, because for ballistic wires with length larger than the decay length of the gapped modes the transmission eigenvalues for the gapless (gapped) modes are $T_n=1$ ($T_n=0$) and the number of gapless Majorana bulk modes (apart from phase transitions) is always even.
Such kind of quantized thermal conductance is generically expected in ballistic  wires in the normal state, but the appearance of quantized thermal conductance in superconducting wires is an exceptional property of this topological phase and it is not accompanied by quantized electric conductance. 

\begin{figure}[b]
\includegraphics[width=1\columnwidth]{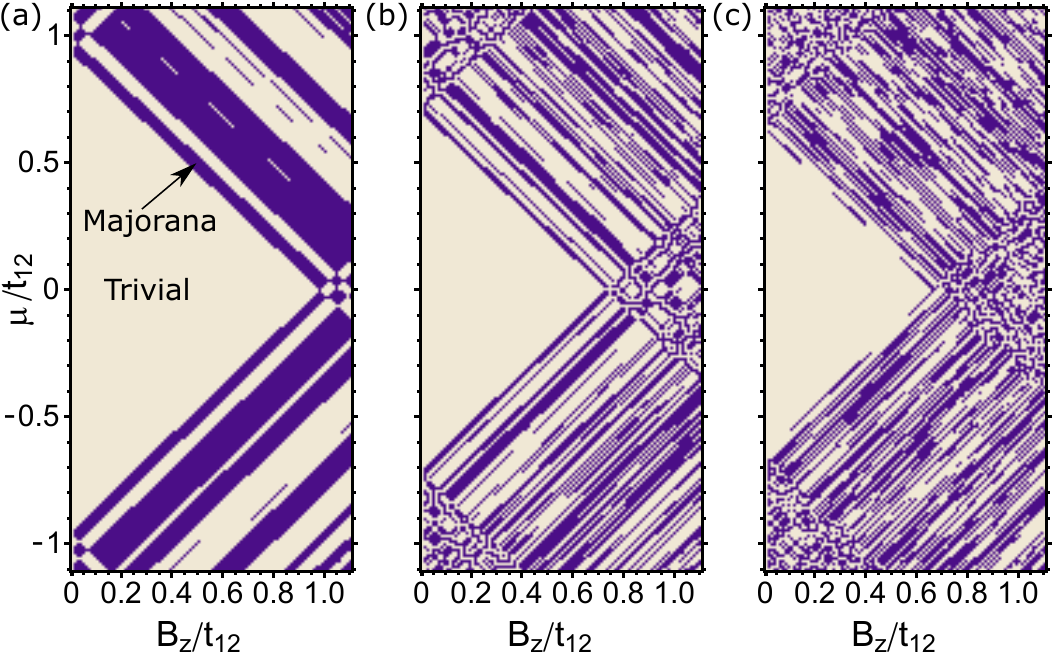}
\caption{Topological phase diagrams for SnTe  nanowires in presence of induced superconductivity. The thicknesses of the nanowires are (a) $8$, (b) $10$ and (c) $18$ atoms. The blue regions indicate parameter regimes where $\nu_{sc}=1$. In the presence of inversion symmetry they correspond to a topological phase supporting gapless Majorana bulk modes. If the inversion symmetry is broken they correspond to fully gapped topological superconducting phase supporting Majorana end modes.
In the numerical calculations we have used $\Delta = 0.01$ eV. The topological region always
starts for $B_z>\Delta$ and the main effect of increasing (decreasing) $\Delta$ is to
shift the nontrivial phases right (left) along $B_z$ axis.  \label{fig:6}}
\end{figure}

\begin{figure}
\includegraphics[width=1\columnwidth]{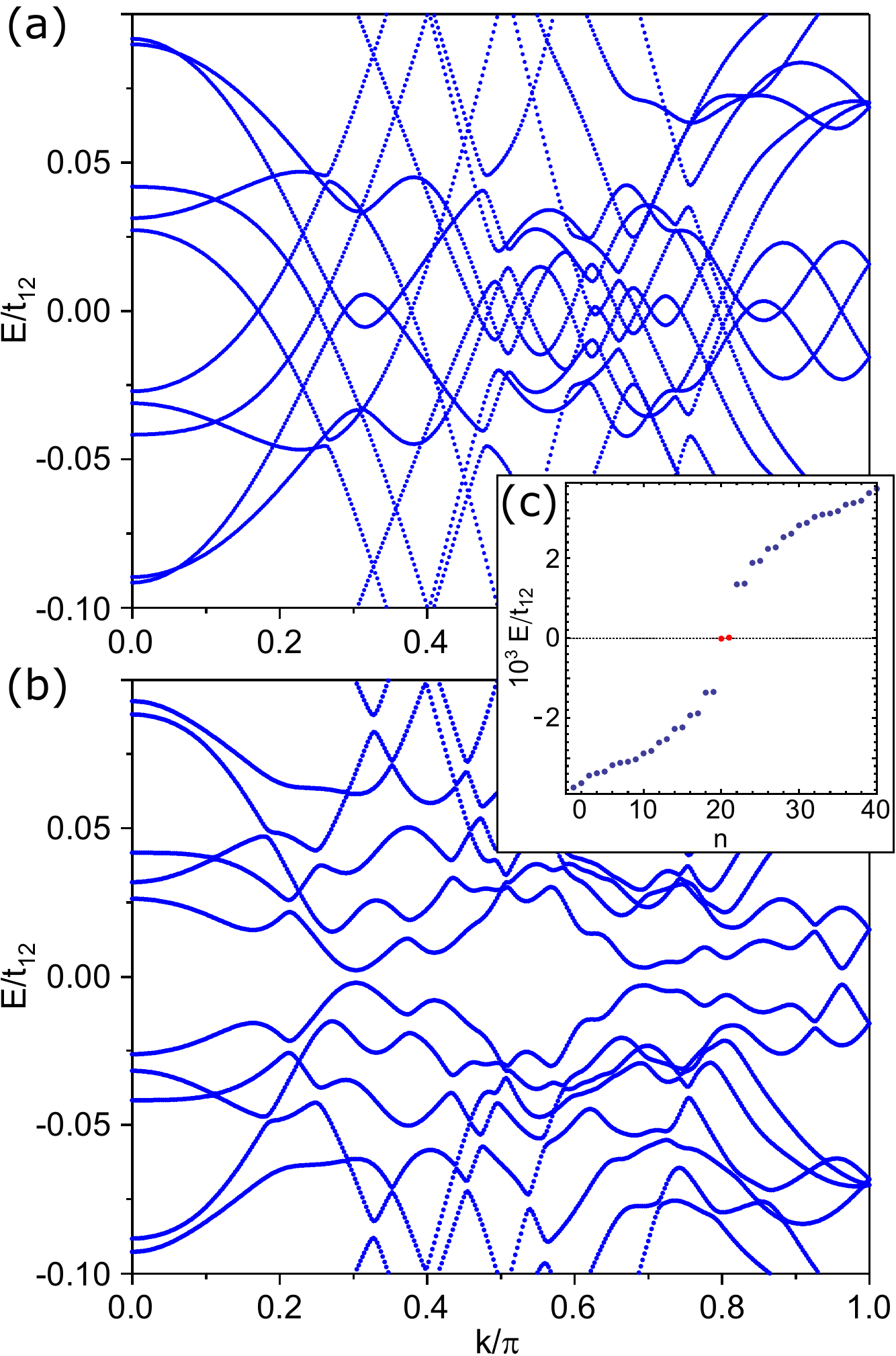}
\caption{Band structures for $8$ atoms thick superconducting nanowires (a) in the presence of inversion symmetry  $\boldsymbol{\lambda}_R=0$ and (b) in the absence of inversion symmetry $\boldsymbol{\lambda}_R=(0,0.05,0)$ eV. (c) The spectrum for 400 atoms long superconducting nanowire, for $\boldsymbol{\lambda}_R=(0,0.05,0)$ eV, demonstrating the existence of zero-energy Majorana modes localized at the end of the wire (red dots). For illustration purposes, we have computed the spectra for very thin nanowires with $B_z=0.16$ eV, $\mu = 0.91$ eV and $\Delta =0.1$ eV. However, due to the general arguments presented in the text, qualitatively similar results are expected also for experimentally feasible values of $B_z$, $\mu$ and $\Delta$ in thicker nanowires.
\label{fig:7}}	
\end{figure}

It turns out that (see Appendix \ref{sec:twopfaffians})
\begin{equation}
{\rm Pf} H^{sc}_U(k_z=0,\pi)={\rm Pf} H_{V}^{sc}(k_z=0,\pi).
\end{equation}
This means that in the presence of inversion symmetry the nontrivial topological invariant $\nu_{sc}=1$ always leads to a change of $\nu_I (k_z)$ between $k_z=0$ and $k_z=\pi$. Thus, in the presence of inversion symmetry there cannot exist fully gapped topologically nontrivial superconducting phase supporting Majorana end modes, but instead $\nu_{sc}=1$ guarantees the existence of topologically nontrivial phase supporting gapless Majorana bulk modes.
To get localized zero-energy Majorana end modes it is necessary to break the inversion symmetry. 

The inversion symmetry can be broken in SnTe nanowires by utilizing ferroelectricity or a structure inversion asymmetry \cite{Chang2016, Kim2019, ZFu2019, Valentine17, Lee2020, Rafal21}. For the results obtained in this paper the explicit mechanism of the inversion symmetry breaking is not important. Therefore, for simplicity we consider a Rashba coupling term  
\begin{equation}
H_R(\mathbf{k})=\boldsymbol{\lambda}_R\cdot\sin\mathbf{k}\times\boldsymbol{\sigma}.
\end{equation}
The magnitude of the Rashba coupling $\boldsymbol{\lambda}_R$ can be considered as the inversion-symmetry breaking field. For $\boldsymbol{\lambda}_R=0$ the inversion symmetry is obeyed and only the gapless topological phase can be realized, whereas for $\boldsymbol{\lambda}_R \ne 0$ the gapless Majorana bulk modes are not protected and the opening of an energy gap can transform the system into a fully gapped topologically nontrivial superconductor supporting localized Majorana end modes.

In Fig.~\ref{fig:6} we illustrate the dependence of the topological phase diagrams on the nanowire thickness. For very small thicknesses there exists a large insulating gap at the charge neutrality point in the normal state spectrum (Fig.~\ref{fig:3}), and therefore the topologically nontrivial phase can be reached only by having either a reasonably large chemical potential $\mu$ or Zeeman field $B_z$. However, with increasing thickness of the nanowire the insulating gap decreases and the nontrivial phases are distributed more uniformly in the parameter space. Similarly as in the case of the normal state phase diagram (Fig.~\ref{fig:2}), we expect that for realistic nanowire thicknesses the topologically nontrivial phase is accessible with experimentally feasible values of the chemical potential and Zeeman field. The structure of the topological phase diagram is quite complicated and it is not easy to extract simple conditions for the existence of the nontrivial phase. It is worth mentioning that the topological region always
starts for $B_z>\Delta$ and the main effect of increasing (decreasing) $\Delta$ is to
shift the nontrivial phases right (left) along $B_z$ axis.

As discussed above, in the presence of inversion symmetry the $\nu_{sc}=1$ regions in Fig.~\ref{fig:6} correspond to the gapless topological phase. This is indeed the case as demonstrated with explicit calculation in Fig.~\ref{fig:7}(a). Moreover, if inversion-symmetry breaking field is introduced our numerical calculations confirm the opening of an energy gap in the bulk spectrum and the appearance of the zero-energy Majorana modes localized at the end of the superconducting nanowire [Fig.~\ref{fig:7}(b),(c)].

\begin{figure}
\includegraphics[width=0.9\columnwidth]{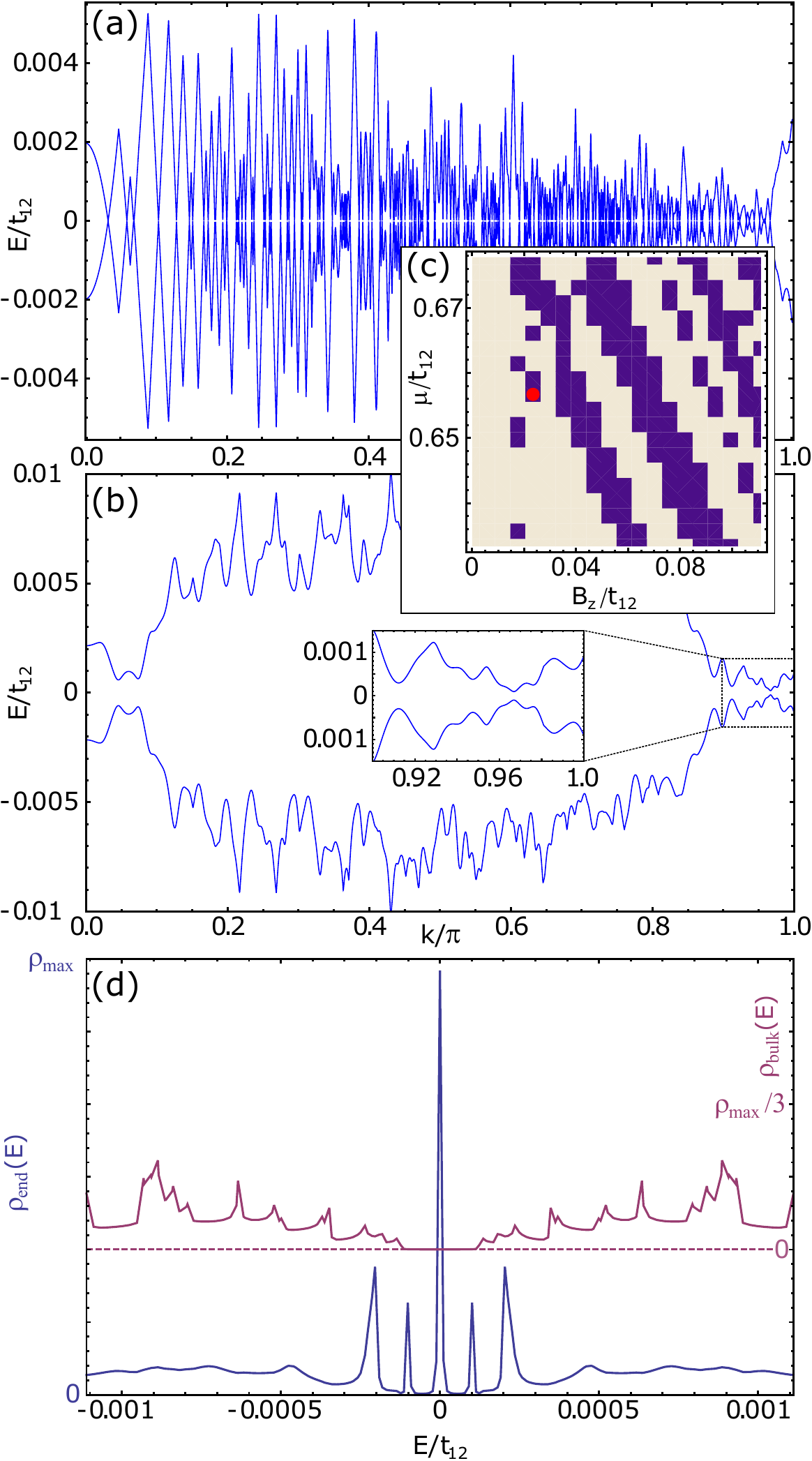}
\caption{Low-energy band structures for $18$ atoms thick superconducting nanowires (a) in the presence of inversion symmetry  $\boldsymbol{\lambda}_R=0$ and (b) in the absence of inversion symmetry $\boldsymbol{\lambda}_R = (0.02, 0.03, 0)$ eV. (c) Zoom into the relevant regime of the phase diagram. We have chosen  $B_z=0.02$ eV, $\mu = 0.5913$ eV and $\Delta = 0.01$ eV (indicated by the red dot in the phase diagram). (d) Local density of states in the bulk $\rho_{{\rm bulk}}(E)$ and at the end $\rho_{{\rm end}}(E)$ (normalized with the maximum value $\rho_{{\rm max}}$) of the wire calculated using the Green function method described in Ref.~\cite{Sancho_1985}.
\label{fig:13}}	
\end{figure}

As illustrated in Fig.~\ref{fig:6} the topological phase becomes fragmented into smaller and smaller regions in the parameter space upon increasing the wire thickness. Therefore, one might be concerned about the experimental feasibility to observe the topological superconductivity in these systems. The systematic analysis of the dependence of the topological gap on the wire thickness goes beyond the scope of this paper, but in Fig.~\ref{fig:13} we have focused on one of the fragmented topological regions in the case of $18$ atoms thick nanowires. Our results show that also in this case it is possible to achieve a topological gap on the order of $1$K and to realize Majorana zero modes at the end of the wire by breaking the inversion symmetry. Therefore, at least in $18$ atoms thick nanowires the observation of the Majorana zero modes is still experimentally feasible.

\section{Discussion and conclusions}

We have shown that SnTe materials support robust corner states and hinge states in the normal state. The topological nature of these states is related to the approximate symmetries of the SnTe nanowires. Some of the approximations, such as the introduction of anisotropic spin-orbit coupling, are quite abstract technical tricks, but they are extremely useful because they allow us to construct well-defined topological invariants. Moreover, we have checked that our approximations are well-controlled and our results are applicable for realistic multivalley nanowires. We have also shown that the higher-order topological invariant, describing the existence of hinge states, is not directly related to the topological crystalline insulator invariant. Therefore, the nontrivial crystalline insulator and higher-order topologies can appear either separately or together. If they appear together the surface states appearing due to topological crystalline insulator phase can coexist with the hinge states appearing due to higher-order topological phase. The higher-order topological invariant is a 2D invariant related to a high-symmetry plane in the momentum space. This plane corresponds to a fixed value of $k_z$ and we have found that both the 2D bulk and the 1D edge are gapped within this plane, so that only the corner states appear. From the practical point of view this means that the surface states arising from the topological crystalline insulator phase and the hinge states arising from the higher-order topological phase are separated in the momentum $k_z$ so that they can coexist in ballistic wires where $k_z$ is a good quantum number.

We have concentrated on relatively thin nanowires. Since the wave functions of the transverse modes transform as a function of the momentum $k_z$ and energy from hinge states to surface states and bulk states, the transverse mode energies do not obey simple parametric dependencies as a function of the nanowire thickness and the Zeeman field. This means that the SnTe nanowires cannot be described by using a low-energy effective $k.p$ theory. This is illustrated in the complicated phase diagram of nanowires, which we have discovered. Nevertheless, from the general trends in the thickness dependence we can extrapolate that for realistic nanowire thicknesses the topologically nontrivial phases can be reached with experimentally feasible values of the Zeeman field. 

Finally, we have found that the superconducting SnTe nanowires support gapless bulk Majorana modes in the presence of inversion symmetry, and by introducing inversion-symmetry-breaking field, the bulk Majorana modes become gapped and topologically protected localized Majorana zero modes appear at the ends of the wire. This finding opens up new possibilities to control and create Majorana zero modes by controlling the inversion-symmetry breaking fields.

There exists various possibilities to experimentally probe the corner states, hinge states and Majorana modes. High-quality transport studies are definitely the best way to study these systems. Ideally, the SnTe bulk materials would be insulators where the Fermi level is inside the insulating gap. The interesting physics, including the topological surface states, hinge states and corner states all appear in this range of energies in the nanowires. Unfortunately, in reality the SnTe bulk materials typically have a large residual carrier density due to defects, which poses a significant obstacle for the studies of topological transport effects. Therefore it is of crucial importance to improve the control of the carrier density in SnTe materials. In comparison to the bulk systems the nanowires have the advantage that the carrier density can be more efficiently controlled with gate voltages.  Tunneling measurements are possible also in the presence of a large carrier density because one can probe the local density of states as a function of energy by voltage-biasing the tip. One may also try to observe the hinge states and corner states using nano-ARPES but obtaining simultaneously both high-spatial and high-energy resolution is a difficult experimental challenge.  The topologically protected gapless Majorana bulk modes could be probed via thermal conductance measurements, and they may also be detectable by measuring electrical shot-noise power or magnetoconductance oscillations in a ring geometry \cite{Akh11}. The Majorana zero modes give rise to various effects, such as a robust zero-bias peak in the conductance \cite{Law09} and $4\pi$ Josephson effect \cite{Kit01,Lut10}, but the ultimate goal in the physics of Majorana zero modes is of course to observe effects directly related to the  non-Abelian Majorana statistics \cite{Sar15, Beenakker20, flensberg2021engineered}.
The Majorana zero modes can be realized even if a significant residual carrier density is present as illustrated in our phase diagrams. However, the new experimental challenge in this case is that the topologically nontrivial phase becomes more and more fragmented in thick wires.

\section*{Acknowledgements}
The work is supported by the Foundation for Polish Science through the IRA Programme
co-financed by EU within SG OP.  W.B. also acknowledges support by Narodowe Centrum Nauki 
(NCN, National Science Centre, Poland) Project No. 2019/34/E/ST3/00404.

\appendix

\section{Construction of the nanowire Hamiltonian and the symmetry operations \label{app:nanowire}}

In this section we give explicit expressions for the different symmetry operators of the nanowire Hamiltonian. Our starting point is the bulk Hamiltonian (\ref{Ham}). The nearest-neighbor hopping matrices are
\begin{equation}
h_x\left(k_{x} \right) = \begin{small}\begin{pmatrix}
0 & 0 & 0& 0 &\gamma^{-}_{k_x} & 0 & 0 & 0 \\
 0 & 0& 0 & 0 & 0 & \gamma^{+}_{k_x} & 0 & 
  0 \\
   0 & 0  & 0 & 0 &  0 & 0 & \gamma^{-}_{k_x}& 0\\
   0& 0& 0& 0& 0& 0& 0&\gamma^{+}_{k_x}\\
   \gamma^{+}_{k_x}& 0& 0&0&0& 0& 
  0 &0 \\
  0& \gamma^{-}_{k_x}& 0& 0& 0& 0& 0& 
  0\\
   0& 0& \gamma^{+}_{k_x}& 0& 0& 0&0& 0\\
    0& 0& 0& \gamma^{-}_{k_x}& 0& 
  0& 0& 0
   \end{pmatrix},\end{small} \label{eqhx} 
\end{equation}
\begin{equation}
h_y\left(k_{y} \right) = \begin{small}\begin{pmatrix}
0 & 0 & 0& 0 &0 & \gamma^{-}_{k_y} & 0 & 0 \\
 0 & 0& 0 & 0 & \gamma^{+}_{k_y} & 0 & 0 & 
  0 \\
   0 & 0  & 0 & 0 &  0 & 0 & 0& \gamma^{+}_{k_y}\\
   0& 0& 0& 0& 0& 0& \gamma^{-}_{k_y}&0\\
0& \gamma^{-}_{k_y}& 0&0&0& 0& 
  0 &0 \\
  \gamma^{+}_{k_y}& 0& 0& 0& 0& 0& 0& 
  0\\
   0& 0& 0& \gamma^{+}_{k_y}& 0& 0& 0& 0\\
    0& 0& \gamma^{-}_{k_y}& 0& 0& 
  0& 0& 0
   \end{pmatrix}\end{small} \label{eqhy} 
\end{equation}
and
\begin{equation}
h_z\left(k_{z} \right) = \begin{small}\begin{pmatrix}
0 & 0 & 0& 0 &0 & 0 & 0 & \gamma^{+}_{k_z} \\
 0 & 0& 0 & 0 & 0 & 0 & \gamma^{+}_{k_z} & 
  0 \\
   0 & 0  & 0 & 0 &  0 & \gamma^{-}_{k_z} & 0& 0\\
   0& 0& 0& 0&  \gamma^{-}_{k_z}&0& 0&0\\
0& 0& 0&\gamma^{+}_{k_z}&0& 0& 
  0 &0 \\
  0& 0&  \gamma^{+}_{k_z}&0& 0& 0& 0& 
  0\\
  0& \gamma^{-}_{k_z}  &0&0& 0& 0& 0& 0\\
     \gamma^{-}_{k_z}&0& 0& 0& 0& 
  0& 0& 0
   \end{pmatrix}\end{small}, \label{eqhz} 
\end{equation}
where $\gamma^{+}_{k_\alpha} = 1+e^{ik_\alpha}$ and  $\gamma^{-}_{k_\alpha} = 1+e^{-ik_\alpha}$. The next-nearest-neighbor hopping matrices are
\begin{equation}
\begin{split}
& h_{xy}\left(k_{x},k_{y} \right) = \\ 
&  \begin{small}\begin{pmatrix}
0 & \phi_{-x,-y}& 0& 0 &0 & 0 & 0 & 0 \\
\phi_{x,y} & 0& 0 & 0 & 0 & 0 & 0 & 
  0 \\
   0 & 0  & 0 & \theta_{-x,y} &  0 & 0 & 0& 0\\
   0& 0& \theta_{x,-y}& 0&  0&0& 0&0\\
  0& 0&  0&0& 0& \theta_{x,-y}& 0& 
  0\\
  0& 0  &0&0& \theta_{-x,y}& 0& 0& 0\\
    0&0& 0& 0& 0& 
  0& 0& \phi_{x,y}\\
  0& 0& 0&0&0& 0& 
  \phi_{-x,-y} &0 \\
   \end{pmatrix}, \end{small} \label{eqhxy} 
   \end{split}
\end{equation}
\begin{equation}
\begin{split}
& h_{yx}\left(k_{x},k_{y} \right) = \\ 
&  \begin{small}\begin{pmatrix}
0 & \theta_{-x,-y}& 0& 0 &0 & 0 & 0 & 0 \\
\theta_{x,y} & 0& 0 & 0 & 0 & 0 & 0 & 
  0 \\
   0 & 0  & 0 & \phi_{-x,y} &  0 & 0 & 0& 0\\
   0& 0& \phi_{x,-y}& 0&  0&0& 0&0\\
  0& 0&  0&0& 0& \phi_{x,-y}& 0& 
  0\\
  0& 0  &0&0& \phi_{-x,y}& 0& 0& 0\\
    0&0& 0& 0& 0& 
  0& 0& \theta_{x,y}\\
  0& 0& 0&0&0& 0& 
  \theta_{-x,-y} &0 \\
   \end{pmatrix},\end{small} \label{eqhyx} 
   \end{split}
\end{equation}
\begin{equation}
\begin{split}
& h_{yz}\left(k_{y},k_{z} \right) = \\ 
& \begin{small}\begin{pmatrix}
0 & 0& \theta_{-y,z}& 0 &0 & 0 & 0 & 0 \\
0 & 0& 0 & \phi_{y,z} & 0 & 0 & 0 & 
  0 \\
   \theta_{y,-z} & 0  & 0 & 0 &  0 & 0 & 0& 0\\
   0& \phi_{-y,-z}& 0& 0&  0&0& 0&0\\
  0& 0&  0&0& 0& 0& \theta_{-y,z}& 
  0\\
  0& 0  &0&0& 0& 0& 0& \phi_{y,z}\\
    0&0& 0& 0& \theta_{y,-z}& 
  0& 0& 0\\
  0& 0& 0&0&0& \phi_{-y,-z}& 
  0 &0 \\
   \end{pmatrix},\end{small} \label{eqhyz} 
   \end{split}
\end{equation}
\begin{equation}
\begin{split}
& h_{zy}\left(k_{y},k_{z} \right) = \\ 
& \begin{small}\begin{pmatrix}
0 & 0& \phi_{-y,z}& 0 &0 & 0 & 0 & 0 \\
0 & 0& 0 & \theta_{y,z} & 0 & 0 & 0 & 
  0 \\
   \phi_{y,-z} & 0  & 0 & 0 &  0 & 0 & 0& 0\\
   0& \theta_{-y,-z}& 0& 0&  0&0& 0&0\\
  0& 0&  0&0& 0& 0& \phi_{-y,z}& 
  0\\
  0& 0  &0&0& 0& 0& 0& \theta_{y,z}\\
    0&0& 0& 0& \phi_{y,-z}& 
  0& 0& 0\\
  0& 0& 0&0&0& \theta_{-y,-z}& 
  0 &0 \\
   \end{pmatrix},\end{small} \label{eqhzy} 
   \end{split}
\end{equation}
\begin{equation}
\begin{split}
& h_{zx}\left(k_{x},k_{z} \right) = \\ 
&  \begin{small}\begin{pmatrix}
0 & 0& 0& \theta_{-x,z} &0 & 0 & 0 & 0 \\
0 & 0 & \phi_{x,z}&0 & 0 & 0 & 0 & 
  0 \\
   0 & \phi_{-x,-z}  & 0 & 0 &  0 & 0 & 0& 0\\
   \theta_{x,-z}& 0& 0& 0&  0&0& 0&0\\
  0& 0&  0&0& 0& 0&0& \phi_{x,z}\\
  0& 0  &0&0& 0& 0&  \theta_{-x,z}&0\\
    0&0& 0& 0&0& \theta_{x,-z}& 
   0& 0\\
  0& 0& 0&0& \phi_{-x,-z}& 0&
  0 &0 \\
   \end{pmatrix} \end{small} \label{eqhzx} 
   \end{split}
\end{equation}
and
\begin{equation}
\begin{split}
& h_{xz}\left(k_{x},k_{z} \right) = \\ 
&  \begin{small}\begin{pmatrix}
0 & 0& 0& \phi_{-x,z} &0 & 0 & 0 & 0 \\
0 & 0 & \theta_{x,z}&0 & 0 & 0 & 0 & 
  0 \\
   0 & \theta_{-x,-z}  & 0 & 0 &  0 & 0 & 0& 0\\
   \phi_{x,-z}& 0& 0& 0&  0&0& 0&0\\
  0& 0&  0&0& 0& 0&0& \theta_{x,z}\\
  0& 0  &0&0& 0& 0&  \phi_{-x,z}&0\\
    0&0& 0& 0&0& \phi_{x,-z}& 
   0& 0\\
  0& 0& 0&0& \theta_{-x,-z}& 0&
  0 &0 \\
   \end{pmatrix},\end{small} \label{eqhxz} 
   \end{split}
\end{equation}\\
where $\theta_{\pm \alpha,\pm \beta} = e^{ \pm i k_\alpha } +e^{ \pm i k_\beta }$ and $\phi_{\pm \alpha,\pm \beta} =1+e^{i\left( \pm k_\alpha \pm k_\beta \right)}$.

To obtain the lower dimensional Hamiltonians, we can first expand the Hamiltonian as
\begin{equation}
\mathcal{H} \left(\textbf{k} \right) = H_{0} \left( k_y,k_z \right)+ e^{-ik_{x}} H_{1} \left( k_y,k_z \right) + e^{ik_{x}} H_{1}^{\dagger} \left( k_y,k_z \right). \nonumber
\end{equation}
Then the 2D Hamiltonian obtained by assuming a finite thickness $N_x$ in $x$-direction is given by $48N_x \times 48 N_x$ matrix
\begin{equation}
\mathcal{H}^{2D}(k_y,k_z)  = \begin{small}\begin{pmatrix}
H_{0} & H_{1}^{\dagger} & 0& 0 &0 & \cdots & 0 \\
 H_{1} & H_{0} & H_{1}^{\dagger} & 0 & 0 &  \cdots & 
  0 \\
   0 &  H_{1} & H_{0} & H_{1}^{\dagger} &  0 &  \cdots & 0\\
   \vdots & \ & \ &\ddots &\ & \ & 
  \vdots &\\
  \vdots & \ & \ & \ & \ddots & \ & \vdots & 
   \\
   0& \cdots & \cdots & 0& 0&H_{0}& H_{1}^{\dagger} \\
    0& \cdots & \cdots & 0& 0& H_{1} & H_{0} 
\end{pmatrix}\end{small}. \label{eqH2D} 
\end{equation}
Similarly, we can decompose this 2D Hamiltonian as
\begin{equation}
\mathcal{H}^{2D}(k_y,k_z)  = H_{0}^{\prime} \left( k_z \right)+ e^{-ik_{y}} H_{1}^{\prime}  \left( k_z \right) + e^{ik_{y}} {H_{1}^{\prime} }^{\dagger} \left( k_z \right),
\end{equation}
where $H_{0 (1)}^{\prime}$ are  $48N_x \times 48 N_x$ matrices. The Hamiltonian for the nanowire with a finite thickness $N_x$ ($N_y$) in $x$ ($y$) direction is given by $48N_x N_y \times 48 N_x N_y$ matrix
\begin{equation}
\mathcal{H}^{1D}(k_z)  = \begin{small}\begin{pmatrix}
H_{0}^{\prime} & {H_{1}^{\prime} }^{\dagger} & 0& 0 &0 & \cdots & 0 \\
 H_{1}^{\prime} & H_{0}^{\prime} & {H_{1}^{\prime} }^{\dagger} & 0 & 0 &  \cdots & 
  0 \\
   0 &  H_{1}^{\prime} & H_{0}^{\prime} & {H_{1}^{\prime} }^{\dagger} &  0 &  \cdots & 0\\
   \vdots & \ & \ &\ddots &\ & \ & 
  \vdots &\\
  \vdots & \ & \ & \ & \ddots & \ & \vdots & 
   \\
   0& \cdots & \cdots & 0& 0&H_{0}^{\prime} & {H_{1}^{\prime} }^{\dagger} \\
    0& \cdots & \cdots & 0& 0& H_{1}^{\prime} &H_{0}^{\prime} 
\end{pmatrix}\end{small}. \label{eqH1D} 
\end{equation}
Assuming that $N_x=N_y$, the nanowire has a 
screw-axis symmetry, which is described by an operator
\begin{equation}
S_c\left(k_{z} \right) = P_z\otimes \exp\left(-i \tfrac{\pi}{4}\sigma_{z}\right)  \otimes \exp\left(-i \tfrac{\pi}{2}L_{z}\right)\otimes  s_c \left( k_z \right). \label{eqSc} 
\end{equation} 
Here $\mathit{P}_z$ is a $N_xN_y \times N_x N_y$ matrix that realizes the fourfold rotation of the unit cells. For general $N_x=N_y$ we have $\mathit{P}_z$ such that $(\mathit{P}_z)_{i,j}=1$ for $i=q+(p-1)N_x$ and $j=p+(N_x-q)N_x$ ($p,q=1,\dots,N_x$) and $(\mathit{P}_z)_{i,j}=0$ otherwise.
$s_c\left(k_{z} \right)$ is the $8 \times 8$ matrix acting inside the unit cell
\begin{equation}
s_c\left(k_{z} \right) = \begin{small}\begin{pmatrix}
0 & 0 & 1& 0 &0 & 0 & 0 & 0 \\
 0 & 0& 0 & 1 & 0 & 0 & 0 & 
  0 \\
   0 & e^{-ik_{z}}  & 0 & 0 &  0 & 0 & 0 & 0\\
   e^{-ik_{z}}& 0& 0&0&0& 0& 
  0 &0 \\
  0& 0& 0& 0& 0& 0& 0& 
  1\\
   0& 0& 0& 0& 0& 0& 1& 0\\
    0& 0& 0& 0& e^{-ik_{z}}& 
  0& 0& 0\\
   0& 0& 0& 0& 0& e^{-ik_{z}}& 0& 0
\end{pmatrix}\end{small}. \label{eqsc} 
\end{equation}
The mirror  symmetry operators corresponding to the mirror planes perpendicular
to $x,y$ and $z$ are
\begin{eqnarray}
\mathit{M}_x(k_{z}) &=&\mathds{1}_{N_y} \!\otimes m_x \otimes  \sigma_x \otimes (2 \mathit{L}_{x}^{2}-\mathds{1}_3 ) \otimes  g(-k_{z}) \mathcal{P}_z \mathcal{P}_x, \nonumber \\ 
\mathit{M}_y (k_{z} ) &=& m_y\otimes \mathds{1}_{N_x} \otimes \sigma_y \otimes (2 \mathit{L}_{y}^{2}-\mathds{1}_3 )\otimes g(-k_{z}) \mathcal{P}_z  \mathcal{P}_y,  \nonumber \\
\mathit{M}_z (k_{z} ) &=& \mathds{1}_{N_y}\otimes \mathds{1}_{N_x}\otimes  \sigma_z  \otimes (2 \mathit{L}_{z}^{2}-\mathds{1}_3 )\otimes g(k_{z}), \label{eqMz}
\end{eqnarray}
where $g\left(k_{z} \right) = {\rm diag} \left(e^{-ik_{z}},e^{-ik_{z}},1,1,e^{-ik_{z}},e^{-ik_{z}},1,1 \right)$,\begin{equation}
\mathcal{P}_x = \begin{small}\begin{pmatrix}
0 & 0 & 0& 0 &1 & 0 & 0 & 0 \\
 0 & 0& 0 & 0 & 0 & 1 & 0 & 
  0 \\
   0 & 0  & 0 & 0 &  0 & 0 & 1 & 0\\
   0& 0& 0&0&0& 0& 
  0 &1 \\
  1& 0& 0& 0& 0& 0& 0& 
 0\\
   0& 1& 0& 0& 0& 0& 0& 0\\
    0& 0& 1& 0& 0& 
  0& 0& 0\\
   0& 0& 0& 1& 0& 0& 0& 0
\end{pmatrix}\end{small} \label{eqPPx} 
\end{equation}
\begin{equation}
\mathcal{P}_y = \begin{small}\begin{pmatrix}
0 & 0 & 0& 0 &0 & 1 & 0 & 0 \\
 0 & 0& 0 & 0 & 1 & 0 & 0 & 
  0 \\
   0 & 0  & 0 & 0 &  0 & 0 & 0 & 1\\
   0& 0& 0&0&0& 0& 
  1 &0 \\
  0& 1& 0& 0& 0& 0& 0& 
 0\\
   1& 0& 0& 0& 0& 0& 0& 0\\
    0& 0& 0& 1& 0& 
  0& 0& 0\\
   0& 0& 1& 0& 0& 0& 0& 0
\end{pmatrix}\end{small} \label{eqPPy} 
\end{equation}
\begin{equation}
\mathcal{P}_z = \begin{small}\begin{pmatrix}
0 & 0 & 0& 0 &0 & 0 & 0 & 1 \\
 0 & 0& 0 & 0 & 0 & 0 & 1 & 
  0 \\
   0 & 0  & 0 & 0 &  0 & 1 & 0 & 0\\
   0& 0& 0&0&1& 0& 
  0 &0 \\
  0& 0& 0& 1& 0& 0& 0& 
 0\\
   0& 0& 1& 0& 0& 0& 0& 0\\
    0& 1& 0& 0& 0& 
  0& 0& 0\\
   1& 0& 0& 0& 0& 0& 0& 0
\end{pmatrix}\end{small} \label{eqPPz} 
\end{equation}
and
\begin{equation}
m_x=m_y = \begin{small}\begin{pmatrix}
0 & \dots & 0 & 1 \\
 \vdots & \ & 1 & 0 \\
   \vdots & \iddots  & \ & \vdots \\
   1& \dots & 0&0
\end{pmatrix}\end{small}. \label{eqmxmy} 
\end{equation}

\section{Weyl semimetal phase for $\lambda_{z}=0$  \label{sec:lz0}}

\begin{figure}[!h]
\includegraphics[width=1\columnwidth]{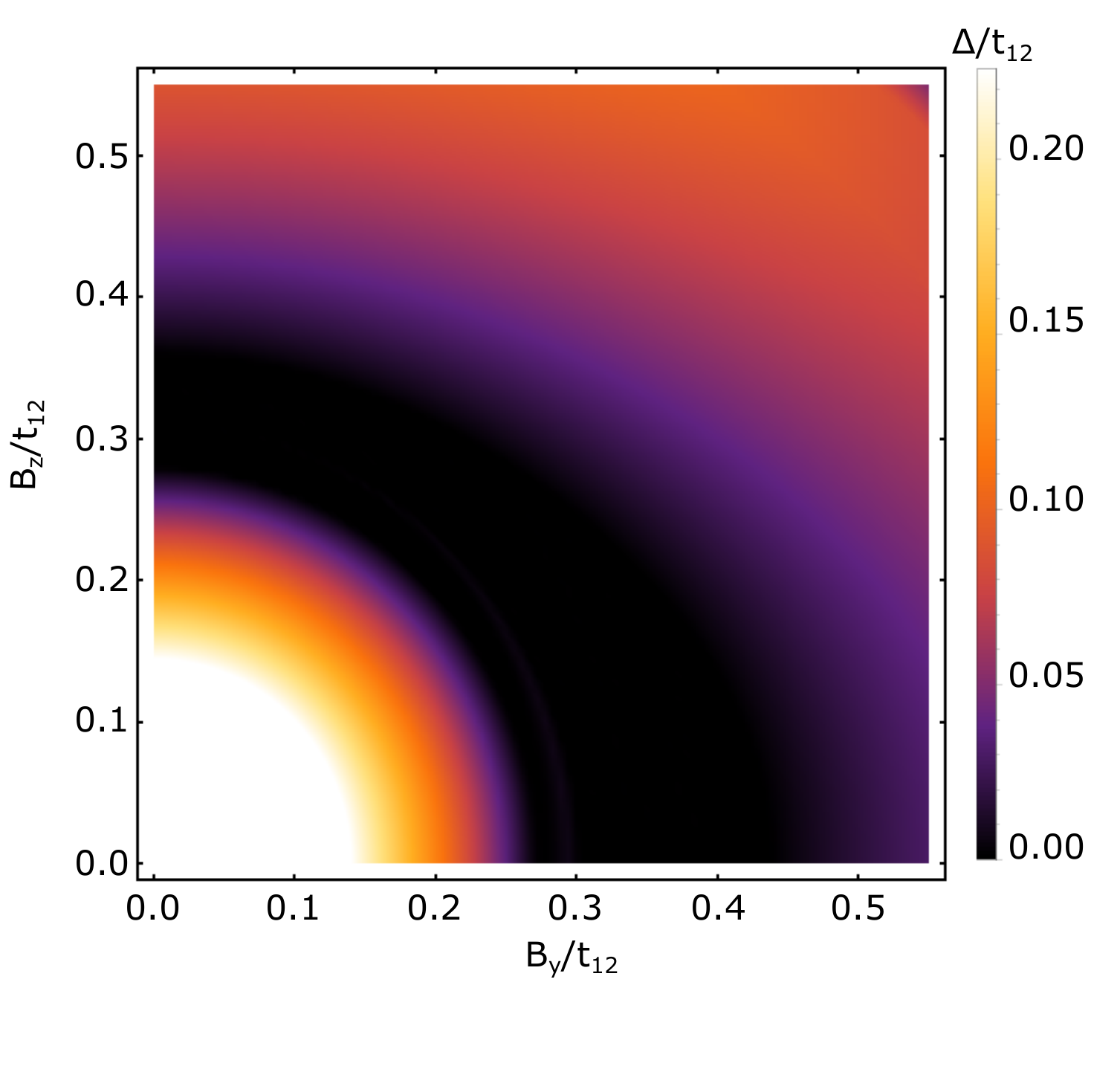}\caption{
The direct band gap $\Delta$ as function of $B_y$ and $B_z$ for 4 atoms thick nanowire and $\lambda_{z}=0$. In the black region $\Delta=0$ and the system is in a 1D Weyl semimetal phase.     \label{fig:app1}}
\end{figure}

Setting $\lambda_{z}=0$ is a good approximation in thin nanowires. In this case, there exists  a non-symmorphic chiral symmetry $S_z(k_z)$ given by, 
\begin{equation}
S_z\left(k_{z} \right) = \mathds{1}_{N_y}\otimes \mathds{1}_{N_x}\otimes\sigma_{z} \otimes \mathds{1}_{3}\otimes \Sigma\mathcal{P}_z g\left( k_z \right),
\end{equation} 
and it is possible to combine it with $M_z(k_z)$  and time-reversal $T$ to construct an antiunitary operator that anticommutes with $\mathcal{H}^{\rm 1D}(k_{z})$ for any $k_z$. This operator squares to $+1$ and therefore it can give rise to a Pfaffian-protected Weyl semimetal phase also when the Zeeman field is not along the $z$-axis.
Fig.~\ref{fig:app1} shows that there exists a Weyl semimetal phase for a range of Zeeman field magnitudes for all Zeeman field directions in the $yz$ plane.

\section{Effects of anisotropic spin-orbit coupling  \label{sec:anisotropy}}

\begin{figure}[!h]
\includegraphics[width=1\columnwidth]{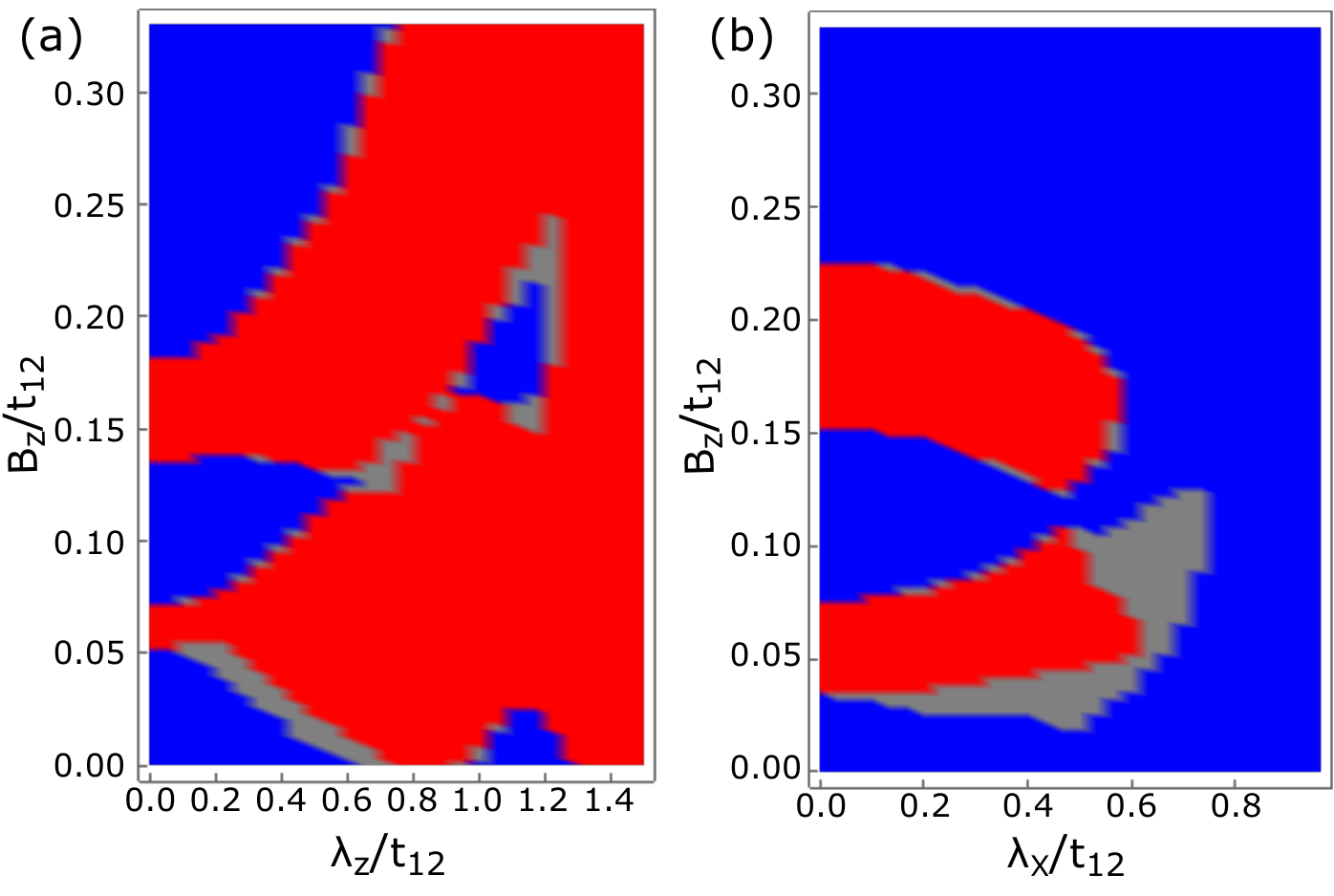}
\caption{ 
Phase diagram as function of (a) $\lambda_z$ and $B_z$ for $\lambda_x=\lambda_y=\lambda$ and (b) $\lambda_x$ and $B_z$ for $\lambda_y=\lambda_z=\lambda$. The different phases are: insulator phase (blue), Weyl semimetal phase (red) and indirect semimetal phase (grey). The thickness of the nanowire is $8$ atoms.    \label{fig:app_anisotropic}}
\end{figure}

We can study the effects of breaking spatial symmetries by considering anisotropic spin-orbit couplings. In Fig.~\ref{fig:app_anisotropic}(a) we show the phase diagram of $8$ atoms thick nanowire as function of $\lambda_z$ and $B_z$ for $\lambda_x=\lambda_y=\lambda$. In this case the system still obeys the screw-axis symmetry so that the phase diagram contains the Weyl semimetal phase in addition to the  insulator and indirect semimetal phases. The phase boundaries depend on $\lambda_z$.

On the other hand, the screw-axis symmetry is broken if $\lambda_x \ne \lambda_y=\lambda_z$. Nevertheless, the system still obeys a two-fold rotational symmetry, which can protect the existence of  Weyl points. Indeed, the phase diagram as a function of $\lambda_x$ and $B_z$ for $\lambda_y=\lambda_z=\lambda$ also contains the Weyl semimetal phase as shown in Fig.~\ref{fig:app_anisotropic}(b).

\section{Higher-order topological invariant}

\subsection{Hamiltonian and symmetries in the $k_3=\pi$ plane \label{sec:2uc}}

In this section we consider the Hamiltonian for a system with atoms at positions $(0,0,\pm1/2)$ and translation vectors $a_1=(1,0,1)$, $a_2=(0,1,1)$
and $(0,0,2)$. For this unit cell the hopping matrices in the bulk Hamiltonian ${\cal H}(\bf{k})$ of Eq.~(\ref{Ham}) are 
\begin{equation}
h_{x}(\bf{k})=\begin{pmatrix}0 & e^{ik_{1}}+e^{i(k_{3}-k_{1})}\\
e^{-ik_{1}}+e^{i(k_{1}-k_{3})} & 0
\end{pmatrix},
\end{equation}
\begin{equation}
h_{y}(\bf{k})=\begin{pmatrix}0 & e^{ik_{2}}+e^{i(k_{3}-k_{2})}\\
e^{-ik_{2}}+e^{i(k_{2}-k_{3})} & 0
\end{pmatrix},
\end{equation}
\begin{equation}
h_{z}(\bf{k})=\begin{pmatrix}0 & 1+e^{ik_{3}}\\
1+e^{-ik_{3}} & 0
\end{pmatrix},
\end{equation}
and
\begin{eqnarray}
h_{xy}(\bf{k}) & = & 2\cos(k_{1}+k_{2}-k_{3})\mathbbm{1}_{2}\nonumber\\
h_{yx}(\bf{k}) & = & 2\cos(k_{1}-k_{2})\mathbbm{1}_{2}\nonumber\\
h_{xz}(\bf{k}) & = & 2\cos(k_{1})\mathbbm{1}_{2}\nonumber\\
h_{zx}(\bf{k}) & = & 2\cos(k_{1}-k_{3})\mathbbm{1}_{2}\nonumber\\
h_{yz}(\bf{k}) & = & 2\cos(k_{2})\mathbbm{1}_{2}\nonumber\\
h_{zy}(\bf{k}) & = & 2\cos(k_{2}-k_{3})\mathbbm{1}_{2},
\end{eqnarray}
and the symmetries discussed above are
\begin{eqnarray}
I&=&\mathbbm{1}_{2}\otimes\mathbbm{1}_{3}\otimes\exp\left( -i\tfrac{k_3}{2}\tau_z\right), \nonumber \\ 
M_{\alpha}&=&\sigma_{\alpha}\otimes(2L_{\alpha}^{2}-\mathbbm{1}_{3})\otimes\mathbbm{1}_{2}, \ \alpha=x,y \nonumber\\
M_{z}&=&\sigma_{z}\otimes(2L_{z}^{2}-\mathbbm{1}_{3})\otimes \exp\left( -i\tfrac{k_3}{2}\tau_z\right) \nonumber \\
M_{xy}
&=&\frac{\sigma_{x}-\sigma_{y}}{\sqrt{2}}\otimes\left(2\left(\frac{L_{x}-L_{y}}{\sqrt 2}\right)^{2}-\mathbbm{1}_{3}\right)\otimes\mathbbm{1}_{2}. \nonumber
\end{eqnarray}
Moreover, we also have a four-fold rotation 
\begin{equation}
R_{4}= \exp\left(i\tfrac{\pi}{2}\tfrac{1}{2}\sigma_{z}\right)\otimes \exp\left(i\tfrac{\pi}{2}L_{z}\right)\otimes \mathbbm{1}_{2}, \nonumber
\end{equation}
and the time-reversal symmetry 
\begin{equation}
T=\sigma_{y}\otimes\mathbbm{1}_{3}\otimes \mathbbm{1}_{2}. \nonumber
\end{equation}
These symmetries act on the Hamiltonian as
\begin{eqnarray}
I{\cal H}(k_{1},k_{2},k_{3})I^{\dagger} & = & {\cal H}(-k_{1},-k_{2},-k_{3})\nonumber\\
M_{x}{\cal H}(k_{1},k_{2},k_{3})M_{x}^{\dagger} & = & {\cal H}(k_{3}-k_{1},k_{2},k_{3})\nonumber\\
M_{y}{\cal H}(k_{1},k_{2},k_{3})M_{y}^{\dagger} & = & {\cal H}(k_{1},k_{3}-k_{2},k_{3})\nonumber\\
M_{z}{\cal H}(k_{1},k_{2},k_{3})M_{z}^{\dagger} & = & {\cal H}(k_{1}-k_{3},k_{2}-k_{3},-k_{3})\nonumber\\
M_{xy}{\cal H}(k_{1},k_{2},k_{3})M_{xy}^{\dagger} & = & {\cal H}(k_{2},k_{1},k_{3})\nonumber\\
R_{4}{\cal H}(k_{1},k_{2},k_{3})R_{4}^{\dagger} & = & {\cal H}(k_{3}-k_{2},k_{1},k_{3}), \nonumber \\
T{\cal H}({\bf k})T^{\dagger}&=&{\cal H}(-{\bf k})^{T}.
\end{eqnarray}
We concentrate on the $k_3=\pi$ plane. In this case, the standard high-symmetry points are related as
\begin{eqnarray}
M_{x}{\cal H}(0,0,\pi)M_{x}^{\dagger} & = & {\cal H}(\pi,0,\pi),\nonumber\\
M_{y}{\cal H}(0,0,\pi)M_{y}^{\dagger} & = & {\cal H}(0,\pi,\pi),\nonumber\\
M_{y}M_{x}{\cal H}(0,0,\pi)M_{x}^{\dagger}M_{y}^{\dagger} & = & {\cal H}(\pi,\pi,\pi).
\end{eqnarray}
Moreover, the $k_3 = \pi$ plane  has four special points 
$(k_1,k_2) = (\pi/2 + n_1\pi,\pi/2 + n_2\pi)$ ($n_{1,2}=0,1$) obeying
\begin{eqnarray}
\left[M_{x},{\cal H}(\tfrac{\pi}{2} + n_1\pi,\tfrac{\pi}{2} + n_2\pi)\right]&=&0 \nonumber\\
\left[M_{y},{\cal H}(\tfrac{\pi}{2} + n_1\pi,\tfrac{\pi}{2} + n_2\pi)\right]&=&0.
\end{eqnarray}
Among these points we find two four-fold rotation centers
\begin{equation}
\left[R_{4},{\cal H}(\tfrac{\pi}{2},\tfrac{\pi}{2},\pi)\right]=\left[R_{4},{\cal H}(\tfrac{3\pi}{2},\tfrac{3\pi}{2},\pi)\right]=0,
\end{equation}
and two four-fold rotoinversion centers
\begin{equation}
\left[Q_{4},{\cal H}(\tfrac{\pi}{2},\tfrac{3\pi}{2},\pi)\right]=\left[Q_{4},{\cal H}(\tfrac{3\pi}{2},\tfrac{\pi}{2},\pi)\right]=0,
\end{equation}
where
\begin{equation}
Q_4 = IR_4.
\end{equation}
The rotoinversion 
centers are mapped onto each other by the four-fold rotation or diagonal mirror
\begin{eqnarray}
R_{4}{\cal H}(\tfrac{\pi}{2},\tfrac{3\pi}{2},\pi)R_{4}^{\dagger}&=&{\cal H}(\tfrac{3\pi}{2},\tfrac{\pi}{2},\pi),\nonumber\\
M_{xy}{\cal H}(\tfrac{\pi}{2},\tfrac{3\pi}{2},\pi)M_{xy}^{\dagger}&=&{\cal H}(\tfrac{3\pi}{2},\tfrac{\pi}{2},\pi),
\end{eqnarray}
and the rotation centers are related by the inversion symmetry 
\begin{equation}
I{\cal H}(\tfrac{\pi}{2},\tfrac{\pi}{2},\pi)I^{\dagger}={\cal H}(\tfrac{3\pi}{2},\tfrac{3\pi}{2},\pi)
\end{equation}

Finally, the product of $M_x$ and $M_y$ yields a two-fold rotation $R_2$ 
with respect to the $z$ axis
\begin{equation}
R_2 = iM_xM_y=\sigma_{z}\otimes(2L_{z}^{2}-\mathbbm{1}_{3})\otimes \mathbbm{1}_2,
\end{equation}
and the rotation and rotoinversion centers are also two-fold rotation centers
\begin{equation}
\left[R_2,{\cal H}(\tfrac{\pi}{2} + n_1\pi,\tfrac{\pi}{2} + n_2\pi)\right]=0, \  n_{1,2}=0,1.
\end{equation} 

\subsection{Pfaffian at the high-symmetry points in the $k_3=\pi$ plane with $\lambda_z=0$ approximation\label{sec:pf}}

By assuming that $\lambda_{z}=0$ and $\lambda_{x}=\lambda_{y}=\lambda$ we find that the Hamiltonian ${\cal H}(k_1,k_2,\pi)$  satisfies a chiral symmetry $S_{z}{\cal H}(k_{1},k_{2},\pi)S_{z}^{\dagger}=-{\cal H}(k_{1},k_{2},\pi)$, where
\begin{equation}
S_{z}=\sigma_{z}\otimes\mathbbm{1}_{3}\otimes\tau_{x}.
\end{equation}
Therefore, the symmetry class of ${\cal H}(k_1,k_2,\pi)$ is 
DIII  so that we can find an eigenbasis of $S_z$ in which
the Hamiltonian takes the form of
\begin{equation}
{\cal H}(k_{1},k_{2},\pi)=\begin{pmatrix}0 & u(k_{1},k_{2})\\
u^{\dagger}(k_{1},k_{2}) & 0
\end{pmatrix},
\end{equation}
and the time-reversal symmetry operator is 
\begin{equation}
T=\begin{pmatrix}0 & -i\mathbbm{1}_{6}\\
i\mathbbm{1}_{6} & 0
\end{pmatrix}.
\end{equation}
Thus, the Hamiltonian satisfies a particle-hole symmetry 
\begin{equation}
C{\cal H}(k_{1},k_{2},\pi)C^{\dagger}=-{\cal H}(-k_{1},-k_{2},\pi)^{T},
\end{equation}
where
\begin{equation}
C=TS_{z}=\begin{pmatrix}0 & \mathbbm{1}_{6}\\
\mathbbm{1}_{6} & 0
\end{pmatrix},
\end{equation}
so that
\begin{equation}
u(k_{1},k_{2})=-u(-k_{1},-k_{2})^{T}. \label{uk}
\end{equation}
Therefore, at the high-symmery points ${\bf K}=(n_1\pi,n_2\pi)$ ($n_{1,2}=0,1$)
\begin{equation}
u(\boldsymbol{K})^{T}=-u(\boldsymbol{K}),
\end{equation}
and we can define a Pfaffian
\begin{equation}
p={\rm Pf}\:u({\bf K}).
\end{equation}
Note that all high-symmetry points $\bf K$ are equivalent. The possible values of $p$ are restricted because of the inversion symmetry operator, which can be written in the present basis as 
\begin{equation}
I=\begin{pmatrix}0 & o\\
o^{T} & 0
\end{pmatrix},
\end{equation}
where
\begin{equation} 
o=\begin{pmatrix}
0 & \mathbbm{1}_{3} \\
-\mathbbm{1}_{3} & 0 
\end{pmatrix}
\end{equation}
is an orthogonal matrix. 
By applying it to the Hamiltonian at the $\bf K$ point we get 
\begin{equation}
u({\bf K})=o u({\bf K})^\dag o  = o u({\bf K})^* o^T,
\end{equation}
where we have used Eq.~(\ref{uk}) and $o^T=-o$. Using the general properties
of the Pfaffian we get 
\begin{equation}
p={\rm Pf}\:u(\boldsymbol{K})= {\rm Pf}\left[ou(\boldsymbol{K})^*o^{T}\right]= 
\det o\:\:{\rm Pf}\:u(\boldsymbol{K})^*=p^{\star}.
\end{equation}
This means that $p$ is a real number.

\subsection{Determinant at the rotation points in the $k_3=\pi$ plane with $\lambda_z=0$ approximation\label{sec:det}}

At the four-fold 
rotation and rotoinversion points we can use these symmetries to decompose the Hamiltonian into diagonal 
blocks. We first focus on the rotoinversion center ${\bf K'}=(\pi/2,3\pi/2,\pi)$ point. In the eigenbasis of $Q_4$ the Hamiltonian takes a block-diagonal form 
\begin{equation}
{\cal H}({\bf K'})=\begin{pmatrix}h_1 & 0 & 0 & 0\\
0 & h_2 & 0 & 0\\
0 & 0 & h_3 & 0\\
0 & 0 & 0 & h_4
\end{pmatrix},
\label{hblock}
\end{equation}
where $h_{1,\dots,4}$ are the $3\times 3$ blocks. By ordering the eigenvalues of $Q_4$ in a suitable way, the chiral symmetry takes a block form
\begin{equation}
S_z=\begin{pmatrix}0 & s_1 & 0 & 0\\
s_2 & 0 & 0 & 0\\
0 & 0 & 0 & s_3\\
0 & 0 & s_4 & 0
\end{pmatrix},
\end{equation}
with $s_{i}$ being $3 \times 3$ unitary block. This form follows from the fact that $Q_4$ and $S_z$ anticommute. 
The spectrum of each individual block $h_i$ is not symmetric around zero, but the spectrum of $h_1$ ($h_3$) is opposite 
to the spectrum of $h_2$  ($h_4$). Since the blocks also have an odd dimension, the determinants satisfy $\det h_1= -\det h_2$ and $\det h_3 = -\det h_4$. 
The spectrum of the whole Hamiltonian ${\cal H}({\bf K'})$ is twice degenerate because of the presence of the symmetry  ${\it \Pi}=IT$
with the property ${\it \Pi}  {\it \Pi}^{\star}=-1$ that gives Kramer denegeracy at every ${\bf k}$ point. 
This symmetry in the present basis takes a block form of 
\begin{equation}
{\it \Pi}=IT=\begin{pmatrix}0 & 0 & 0 & k_1\\
0 & 0 & k_2 & 0\\
0 & k_2^{\dagger} & 0 & 0\\
k_1^{\dagger} & 0 & 0 & 0
\end{pmatrix}.
\end{equation}
This implies that the blocks $h_1$ and $h_4$ ($h_2$ and $h_3$) have the same spectrum. From this it follows that 
\begin{equation}
{\rm det}[{\cal H}({\bf K'})] =d^4,
\end{equation}
where $d \equiv \det h_1=\det h_4 = - \det h_2 = - \det h_3$, 
and therefore $d$ changes sign at the zero-energy gap closing occurring at the momentum ${\bf K'}$. 

Finally, it is worth noticing that the above construction does not work for the four-fold rotation points. In the eigenbasis of $R_4$ the Hamiltonian consists of four diagonal blocks $h_{1,\dots,4}$, where
 $h_{1,2}$ ($h_{3,4}$) are $2 \times 2$ ($4\times 4$) matrices. The rotation $R_4$ commutes 
with $S_z$, so that in this basis
\begin{equation}
S_z=\begin{pmatrix}s_1 & 0 & 0 & 0\\
0 & s_2 & 0 & 0\\
0 & 0 & s_3 & 0\\
0 & 0 & 0 & s_4
\end{pmatrix}.
\end{equation}
From this structure it follows that $\{h_i,s_i\}=0$ so  that spectrum of each block $h_i$ is 
symmetric around zero. Thus, the determinants always satisfy $\det h_{1,2}\le 0$  and $\det h_{3,4}\ge 0$, and therefore they cannot change sign in a gap closing.

\section{Topological invariants in the presence of superconductivity}

\subsection{Topological invariant of 1D superconductors belonging to class D \label{sec:strongsc}}

In the presence of induced superconductivity the BdG Hamiltonian $H^{sc}(k_z)$ always satisfies a particle-hole symmetry $C^{sc} (H^{sc}(-k_z))^T C^{sc} = - H^{sc}(k_z)$, where $C^{sc}$ can be written in the Nambu space as
\begin{equation}
C^{sc}=\begin{pmatrix} 0 & \mathds{1} \\ \mathds{1} & 0\end{pmatrix}.
\end{equation}
We can utilize $C^{sc}$ to perform a unitary transformation on the Hamiltonian
\begin{equation}
H_{U}^{sc}(k_z)=U^{\dagger}H^{sc}(k_z)U, \ U=\beta\sqrt{\Lambda_{C}}^{-1},
\end{equation}
where the columns of matrix $\beta$ are the eigenvectors of $C^{sc}$ and $\Lambda_{C}$
is a diagonal matrix containing the eigenvalues of $C^{sc}$, so that
\begin{equation}
C^{sc}\beta=\beta\Lambda_{C}.
\end{equation}
Because $C^{sc}C^{sc*}=1$, which follows from the D symmetry class, and $C^{sc}$ is unitary we
can choose the eigenvectors so that they satisfy $\beta = \beta^*$. From this it follows that at $k_{0}=0,\pi$
\begin{eqnarray}
\left(H_{U}^{sc}(k_0)\right)^{T}  &=&  \sqrt{\Lambda_C}^{-1} \beta^T (H^{sc}(k_0))^T \beta \sqrt{\Lambda_{C}} \label{fin}\\
&=& \sqrt{\Lambda_C} \beta^T C^{sc} (H^{sc}(k_0))^T C^{sc} \beta \sqrt{\Lambda_{C}}^{-1} \nonumber\\ 
&=&  - \sqrt{\Lambda_C} \beta^T  H^{sc}(k_0) \beta \sqrt{\Lambda_{C}}^{-1}=-H_{U}^{sc}(k_0) \nonumber
\end{eqnarray}
and the particle-hole operator in the new basis becomes identity matrix
\begin{equation}
U^{T}C^{sc}U = \sqrt{\Lambda_C}^{-1} \beta^T C^{sc}  \beta\sqrt{\Lambda_{C}}^{-1}=1.
\end{equation}

Since the Hamiltonian is antisymmetric at $k_0=0,\pi$ we can define a Pfaffian, which is real because
\begin{equation}
[{\rm Pf} H^{sc}_U(k_0)]^*={\rm Pf} H^{sc}_U(k_0)^T={\rm Pf}[H^{sc}_U(k_0)].
\end{equation}
Therefore we can define a $\mathbb{Z}_2$ topological invariant as 
\begin{equation}
\nu_{sc}=\left(1-{\rm sgn}[{\rm Pf} H^{sc}_U(0)  {\rm Pf}  H^{sc}_U(\pi)] \right)/2.
\end{equation}
This is the strong topological invariant of 1D superconductors belonging to the class D.

\subsection{Topological invariant for inversion-symmetry protected gapless Majorana bulk modes \label{sec:inv-bulkMajo}}

We can also combine $C^{sc}$ with the inversion symmetry $I^{sc}(k_z)$ to produce an operator 
\begin{equation}
A^{sc}(k_z)=C^{sc}I^{sc}(k_z).
\end{equation}
 whose action on the Hamiltonian is 
 \begin{equation}
 A^{sc\dagger}_{k_z} \left(H^{sc}(k_z)\right)^{T} A^{sc}_{k_z}=-H^{sc}(k_z).
 \end{equation}
From the double application of the above equation it follows that 
$A^{sc}A^{sc*}=\pm 1$, where $+1$ and $-1$ define different symmetry classes,
in analogy to the particle hole symmetry. In our case we get  $A^{sc}A^{sc*}=+1$,
following from the D symmetry classs, $[C^{sc},I^{sc}]=0$ and $I^{sc}I^{sc*}=1$.

Note that it is enough to know that $A^{sc}$ is unitary and $A^{sc}A^{sc*}=+1$ to
prove that it can be diagonalized by an orthogonal transformation. From unitarity we have
$A^{sc}=\exp( i B)$ with $B$ being Hermitian. From the latter property we get 
$\exp(- i B) =\exp( -i B^T)$, which gives $B^T = B +2n\pi$. By taking trace of this equation
we get that $n=0$ so $B$ must be real symmetric. Then  $A^{sc}$ can be diagonalized by 
by an orthogonal transformation.  Then we find the real eigenbasis $\gamma(k_z)$ of
$A^{sc}(k_z)$, we have:
\begin{equation}
A^{sc}(k_z)\gamma(k_z)=\gamma(k_z)\Lambda_{A}(k_z).
\end{equation}
We define a unitary transformation
\begin{equation}
V(k_z)=\gamma(k_z)\sqrt{\Lambda_{A}(k_z)}^{-1},
\end{equation}
and following the same derivation as in Eq. (\ref{fin}) we 
can prove that the transformed Hamiltonian
\begin{equation}
H_{V}^{sc}(k_z)=V(k_z)^{\dagger}H^{sc} (k_z)V(k_z), 
\end{equation}
is antisymmetric for any $k_z$. 
By utilizing the fact that the Pfaffian of Hamiltonian is real valued, 
we can now define an inversion-symmetry protected $\mathbb{Z}_2$ topological invariant for all values of $k_z$ as
\begin{equation}
\nu_I (k_z) = \left(1- {\rm sgn}[{\rm Pf} H_{V}^{sc}(k_z)] \right)/2.
\end{equation}
If this invariant changes as a function of $k_z$ there must necessarily be a gap closing. Therefore, there exists a 1D topological phase supporting inversion-symmetry protected gapless bulk Majorana modes. In the presence of inversion symmetry these gapless Majorana bulk modes can only be destroyed by merging them in a pairwise manner.

\subsection{Relationship between the two Pfaffians \label{sec:twopfaffians}}

We have two antisymmetric forms of Hamiltonian. $H_{U}^{sc}(k_z)$
is antisymmetric at $k_z=0,\pi$ and $H_{V}^{sc}(k_z)$
is antisymmetric for all values of $k_z$. These antisymmetric Hamiltonians allow us to define topological invariants with the help of their Pfaffians, and thus it is important to know how these Pfaffians are related to each other. Assume we have two antisymmetric Hamiltonians $H'$ and
$H$ related by a change of basis as $H'=Q^{\dagger}HQ$. Then from
antisymmetry of $H'$ and $H$ we have that $\left[H,QQ^{T}\right]=0$.
Then we have two options, either $QQ^{T}=1$ and then ${\rm Pf}H'=\det Q\:{\rm Pf}H$
or $QQ^{T}$ is equal to a symmetry operator of $H$ and then the
Pfaffians of $H$ and $H'$ do not have to be proportional. The latter
case could lead to two independent invariants.

Coming back to our case, we find that 
\begin{equation}
V(k_z=0,\pi)V(k_z=0,\pi)^{T}\not=1,
\end{equation}
and this operator is indeed related to the inversion symmetry
\begin{equation}
V(k_z=0,\pi)V(k_z=0,\pi)^{T} = I^{sc}(k_z=0,\pi),
\end{equation}
but it turns out that also a stronger property holds, namely
\begin{equation}
\left(V(k_z)V(k_z)^{T}\right)^{\star} =  I^{sc}(k_z).
\end{equation}
From the last equation we obtain
\begin{equation}
V(k_z)^{\dagger}=V(k_z)^{T}I^{sc}(k_z).
\end{equation}
Consequently,
\begin{equation}
H_{V}^{sc}(k_z)=V(k_z)^{T} I^{sc}(k_z)H_{U}^{sc}(k_z) V(k_z).
\end{equation}
Thus the Pfaffians can be related as 
\begin{equation}
{\rm Pf}H_{V}^{sc}(k_z)=\det V(k_z)\,{\rm Pf}\left[I^{sc} (k_z) H_U^{sc} (k_z)\right].
\end{equation}
Both sides of the equations are well defined because $H_{V}^{sc}(k_z)$
is antisymmetric for any $k_z$ and $I^{sc} (k_z) H_{U}^{sc}(k_z)$
is also antisymmetric for any $k_z$ despite $H_{U}^{sc}(k_z)$ alone being
symmetric only at high-symmetry points. For the right-hand side we
get $\det V(k_z)=\exp\left[24iN_{x}N_{y}k_z\right]$ so we always have
$\det V(0)=\det V(\pi)=1$. To calculate the Pfaffian on the right-hand side at
 $k_0 = 0, \pi$ we can utilize the operator $\gamma(k_0)=\gamma(k_0)^*$, satisfying $\det\gamma(k_0)=1$ and $\gamma(k_0)^T I^{sc}(k_0) \gamma(k_0)$ is a diagonal matrix, to obtain
 \begin{widetext}
\begin{eqnarray*}
{\rm Pf}H_{V}^{sc}(k_0) &=& {\rm Pf}\left[\gamma(k_0)^{T}I^{sc}(k_0)\gamma(k_0)\gamma(k_0)^{T}H_{U}^{sc}(k_0)\gamma(k_0)\right] = {\rm Pf}\left[-\gamma_{-}(k_0)^{T}H_{U}^{sc}(k_0)\gamma_{-}(k_0)\right]{\rm Pf}\left[\gamma_{+}(k_0)^{T}H_{U}^{sc}(k_0)\gamma_{+}(k_0)\right]\\
	&=& \left(-1\right)^{d_{-}/2}{\rm Pf}\left[\gamma_{-}(k_0)^{T}H_{U}^{sc}(k_0)\gamma_{-}(k_0)\right]{\rm Pf}\left[\gamma_{+}(k_0)^{T}H_{U}^{sc}(k_0)\gamma_{+}(k_0)\right] = {\rm Pf}H_{U}^{sc}(k_0).
\end{eqnarray*}
\end{widetext}
Here, $\gamma(k_0)= [\gamma_+(k_0), \gamma_-(k_0)]$, the columns of $\gamma_\pm(k_0)$  are the eigenvectors of $I^{sc}(k_0)$ corresponding to eigenvalues $\pm 1$ and we have utilized the fact that the dimension $d_-$  of the eigenvalue $-1$ subspace is always a multiple of $4$.  Therefore, the two Pfaffians are always equal at the high-symmetry momenta $k_0=0,\pi$.

\bibliography{refNWs}

\end{document}